\documentclass[12pt]{article}
\usepackage{lipsum}

\usepackage{filecontents}

\usepackage{amsthm}
\usepackage{fullpage}
\usepackage{footnote}
\usepackage{tabu}
\usepackage{booktabs}
\usepackage[title]{appendix}
\usepackage{scribe}
\usepackage{caption}
\usepackage{setspace}
\providecommand{\keywords}[1]
{
  \small	
  \textbf{\textit{Keywords---}} #1
}
\usepackage{etoolbox}

\makeatletter
\patchcmd{\@makecaption}
  {\parbox}
  {\advance\@tempdima-\fontdimen2} % decrease the width!
  {}{}
\makeatother 

\usepackage{titling} 

\usepackage{amsmath,bbm,amssymb, amsfonts}
\usepackage{mathtools}
\usepackage{graphicx,psfrag,epsf}
\usepackage{enumitem}
\usepackage{natbib}
\usepackage{url} % not crucial - just used below for the URL 
\usepackage{color}
\usepackage{algorithmicx}
\usepackage{booktabs}
\usepackage{siunitx}
\usepackage{array}
\usepackage{rotating}
\usepackage{adjustbox}

\usepackage{algorithm}
\usepackage{algpseudocode}

\usepackage{tikz}

\usepackage{url}

\usepackage{multicol}
\usepackage{xcolor}
\usepackage{algorithm}
\makeatletter
\def\breve{\mathpalette\wide@breve}
\def\wide@breve#1#2{\sbox\z@{$#1#2$}%
	\mathop{\vbox{\m@th\ialign{##\crcr
				\kern0.08em\brevefill#1{0.8\wd\z@}\crcr\noalign{\nointerlineskip}%
				$\hss#1#2\hss$\crcr}}}\limits}
\def\brevefill#1#2{$\m@th\sbox\tw@{$#1($}%
	\hss\resizebox{#2}{\wd\tw@}{\rotatebox[origin=c]{90}{\upshape(}}\hss$}
\makeatletter

\newtheorem{theorem}{Theorem}

\def\gobblestop#1#2{#1}
\def\killstop{%
	\aftergroup\gobblestop
}
\def\thick#1{\hbox{\rlap{$#1$}\kern0.25pt\rlap{$#1$}\kern0.25pt$#1$}}

\def\smbalpha{\boldsymbol{{\scriptstyle{\alpha}}}}

\def\smbalpha{\widehat{\smbalpha}}

\def\hbar{\bar{ h}}

\def\mybox#1{\vskip1mm \begin{center}
        \hspace{.0\textwidth}\vbox{\hrule\hbox{\vrule\kern6pt
\parbox{.9\textwidth}{\kern6pt#1\vskip6pt}\kern6pt\vrule}\hrule}
        \end{center} \vskip-5mm}
\def\lboxit#1{\vbox{\hrule\hbox{\vrule\kern6pt
      \vbox{\kern6pt#1\vskip6pt}\kern6pt\vrule}\hrule}}

\def\thickboxit#1{\vbox{{\hrule height 1mm}\hbox{{\vrule width 1mm}\kern6pt
          \vbox{\kern6pt#1\kern6pt}\kern6pt{\vrule width 1mm}}
               {\hrule height 1mm}}}

\def\fat#1{\hbox{\rlap{$#1$}\kern0.25pt\rlap{$#1$}\kern0.25pt$#1$}}

\newcolumntype{R}{@{\extracolsep{0.5cm}}r@{\extracolsep{0pt}}}%            
\newcolumntype{E}{@{\extracolsep{0.25cm}}c@{\extracolsep{0pt}}}%

\makeatletter
\newcommand{\distas}[1]{\mathbin{\overset{#1}{\kern\z@\sim}}}%
\makeatother

\usepackage{xr-hyper}
\usepackage[colorlinks, linkcolor=blue, anchorcolor=blue, citecolor=blue]{hyperref}

\newtheorem{prop}[theorem]{Proposition}

\newtheorem{lemma}[theorem]{Lemma}

\usepackage[left=1in,right=1in,top=1in,bottom=1in]{geometry}

\usepackage{xr-hyper}
\makeatletter
\newcommand*{\addFileDependency}[1]{% argument=file name and extension
  \typeout{(#1)}
  \@addtofilelist{#1}
  \IfFileExists{#1}{}{\typeout{No file #1.}}
}
\makeatother

\newcommand{\blind}{1}

\begin{document}

\if1\blind
{
	\title{\bf Exploring the effects of mechanical ventilator settings with modified vector-valued treatment policies}
	\author{Ziren Jiang$^{1}$, Philip S. Crooke$^{2}$, John J. Marini$^{3}$ Jared D. Huling$^{1}$\thanks{corresponding author: huling@umn.edu}\\ [8pt]
		$^{1}$Division of Biostatistics and Health Data Science,\\ University of Minnesota \\ [4pt]
        $^{2}$Department of Mathematics, Vanderbilt University \\ [4pt]
        $^{3}$Department of Pulmonary and Critical Care Medicine, \\ University of Minnesota  \\ [8pt]
	}
        \date{}
	\maketitle
} \fi

\if0\blind
{
	\title{\bf Exploring the effects of mechanical ventilator settings with modified vector-valued treatment policies}
	\date{}
	\maketitle
} \fi

%\pagenumbering{gobble}

\begin{abstract}
Mechanical ventilation is critical for managing respiratory failure, but inappropriate ventilator settings can lead to ventilator-induced lung injury (VILI), increasing patient morbidity and mortality. Evaluating the causal impact of ventilator settings is challenging due to the complex interplay of multiple treatment variables and strong confounding due to ventilator guidelines. In this paper, we propose a modified vector-valued treatment policy (MVTP) framework coupled with energy balancing weights to estimate causal effects involving multiple continuous ventilator parameters simultaneously in addition to sensitivity analysis to unmeasured confounding. Our approach mitigates common challenges in causal inference for vector-valued treatments, such as infeasible treatment combinations, stringent positivity assumptions, and interpretability concerns. Using the MIMIC-III database, our analyses suggest that equal reductions in the total power of ventilation (i.e., the mechanical power) through different ventilator parameters result in different expected patient outcomes. Specifically, lowering airway pressures may yield greater reductions in patient mortality compared to proportional adjustments of tidal volume alone. Moreover, controlling for respiratory-system compliance and minute ventilation, we found a significant benefit of reducing driving pressure in patients with acute respiratory distress syndrome (ARDS). Our analyses help shed light on the contributors to VILI.
 \\[3em]    
\end{abstract}%

\newpage
%\spacingset{1.75} % DON'T change the spacing!
\setstretch{1.775}
%\spacingset{1.5} % DON'T change the spacing!
\setlength{\abovedisplayskip}{7pt}%
\setlength{\belowdisplayskip}{7pt}%
\setlength{\abovedisplayshortskip}{5pt}%
\setlength{\belowdisplayshortskip}{5pt}%
%\pagenumbering{arabic}
%\setcounter{page}{1}
%%%%   ------------------------------------
%%%%     Input introduction section
%%%%   ------------------------------------
\section{Introduction}\label{sec:1}

Mechanical ventilation is a lifesaving intervention for patients with respiratory failure, providing critical support to maintain adequate gas exchange. However, the application of mechanical ventilation is not without risks. With each breath delivered by a mechanical ventilator, a specific amount of energy is transferred to the respiratory system, primarily to overcome airway resistance and expand lung tissue. Only a portion of this energy is recovered through the elastic recoil of the lung, while the remainder is dissipated in the lung parenchyma \citep{busana2022energy}. If ventilator settings are not carefully optimized, this process may result in ventilator-induced lung injury (VILI), a phenomenon believed linked to poor outcomes, including increased morbidity and mortality \citep{huhle2018mechanical}. Understanding how the various settings impact mortality is thus of great interest.

The process of one ventilation cycle in terms of the airway pressure is illustrated in the left panel of Figure \ref{fig:MPintro}. Each cycle can be characterized by the following ventilator settings: (1) tidal volume (${V_T}$), which represents the amount of air delivered per ventilation cycle; (2) respiratory rate (${{RR}}$), which is the frequency at which ventilation cycles occur per minute; (3) peak inspiratory pressure (${P_{\textnormal{peak}}}$), which is the maximum pressure applied to the lungs during inhalation (4) plateau pressure (${P_{\textnormal{plateau}}}$), which is the static airway pressure when airflow is paused at the end of inhalation; and (5) positive end-expiratory pressure ($\textnormal{{PEEP}}$), which is the baseline pressure maintained at the end of exhalation to prevent alveolar collapse. 

The possibility of VILI has long been recognized in clinical practice, prompting extensive research aimed at understanding its underlying mechanisms and developing lung-protective ventilation strategies to mitigate this complication. Pivotal randomized controlled trials (RCTs) showed that mechanical ventilation with low tidal volumes (${V_T}$) improves patient survival \citep{amato1998effect,acute2000ventilation}. Various variables \textit{alone} have been demonstrated experimentally to associate with VILI, including respiratory rate ($\textnormal{RR}$) \citep{rich2003effect}, and driving pressure (defined as $\textnormal{DP} = P_{\textnormal{plateau}} - \textnormal{PEEP}$) \citep{bellani2011lung}. However, each setting contributes to a different aspect of ventilation  and their combination is what results in a particular ventilation cycle. Under passive conditions, the product of energy per cycle and RR per minute has become known in medical practice as `power'. Mechanical power ($\textnormal{MP}$) is defined as:
\begin{equation}\label{eqn:mp}
    \textnormal{MP} = 0.098\times V_T \times \textnormal{RR} \times [P_{\textnormal{peak}}-\frac{1}{2}\times (P_{\textnormal{plateau}}-\textnormal{PEEP})],
\end{equation}
which is a putative summary metric that aggregates the contribution of the ventilator settings to VILI \citep{neto2018mechanical, vasques2018mechanical}. Reasoning that energy is required to accomplish alveolar stretch, the rate of mechanical energy delivery per minute has been viewed as a unifying metric to quantify VILI risk—whatever its components.

Literature has shown that the larger values of MP are associated with higher mortality \citep{neto2018mechanical}, indicating that indeed ventilator settings may cause harm to patients. However, as a summary measurement, MP is not directly manipulable, making a single causal relationship between MP and mortality ill-posed. 
Further, \citet{marini2020component} hypothesized that it `\textit{seems unlikely that all combinations of frequency, tidal volume, and pressure (flow resistive pressure, driving pressure, and PEEP) that sum to the same power value are equally dangerous to a given lung}.' Providing sharper insights into the different contributors of VILI are thus essential for providing better ventilation recommendations.

However, existing literature with observational studies has primarily focused on a {single} ventilation variable at each time, which may lead to biased conclusions because settings are not set in isolation from each other in practice. Only targeting one of the settings at a time, manipulating one setting among the many that exist may intrinsically change the value of other settings, thus entangling the interpretation of the causal effect. For example, ${P_{\textnormal{plateau}}}$ and ${P_{\textnormal{peak}}}$ both measure respiratory system pressures, though at different ventilation stages, and exhibit a strong empirical correlation of 0.86 in our motivating dataset. Thus, assessment of the causal effect of reducing ${P_{\textnormal{peak}}}$ alone implicitly involves comparing patients who have lower ${P_{\textnormal{plateau}}}$. Methods that explicitly allow for joint manipulation of treatments are thus needed to provide clearer insights into the causes of VILI.

In this work, we analyze the MIMIC-III (Medical Information Mart for Intensive Care) dataset \citet{johnson2016mimic} with simultaneous consideration of multiple variables for mechanical ventilation in exploring their relationship to ICU patient mortality. We use MIMIC-III in particular, as the time-frame during which data were collected was prior to a shift in ventilator use due to guideline changes in part as a result of the analyses of \citet{amato2015driving} of data from several randomized trials. Post 2015, variation in ventilation use is likely less \citep{del2017mechanical}, making positivity violations even more severe and thus causal analyses more challenging.

\subsection{Causal inference for vector-valued treatments}
Consider data $\{\boldsymbol{X}_i,\A_i,Y_i\}_{i=1}^n$ collected from an observational study, where for subject $i=1,...,n$, $\X_i \in \mathcal{X} \subseteq\mathbb{R}^p$ denotes a $p$-dimensional vector of pre-treatment covariates, $\A_i \in \mathcal{A} \subseteq \mathbb{R}^k$ denotes the vector of $k$ real-valued intervention variables, and $Y_i\in \mathbb{R}$ denotes the outcome.  We are interested in assessing the causal effects of manipulating multiple treatment variables (exposures) simultaneously. While this work focuses on mechanical ventilation, our methods have utility in far broader areas of research, including the impact of health policy on childhood obesity \citep{williams2020causal}, environmental health considering the joint effects of different pollutants \citep{zhu2020joint} and many others \citep{wages2011dose}.

Several approaches have been proposed for estimating causal effects involving vector-valued treatments. \cite{egger2013generalized} extended the generalized propensity score (GPS) framework to scenarios involving multiple continuous treatments and demonstrated its effectiveness in finite samples. \cite{nabi2022semiparametric} introduced a semiparametric dimension-reduction method specifically tailored for high-dimensional treatment settings, explicitly considering the magnitude of causal relationships. For situations with multiple binary/categorical treatment(s), \citet{li2019propensity} proposed using balancing weights derived from generalized propensity scores to balance the covariate distribution to a prespecified target population. They also developed generalized overlap weights suited for these categorical treatment scenarios. \citet{pashley2023causal} addressed multiple binary treatments by conceptualizing observational studies as hypothetical fractional factorial experiments. Despite these advances, causal inference involving vector-valued treatments remains challenging, as unique complexities inherent in these scenarios are not fully addressed by existing methods.

A key difficulty in causal inference with vector-valued treatments is the fragility of the positivity assumption, primarily due to two factors. From a scientific perspective, not all combinations of treatment values are feasible or clinically appropriate for every individual. Consequently, assuming a positive probability for all treatment combinations among all participants, as in many existing literature, is unrealistic. For instance, setting both respiratory rate (RR) and tidal volume (TV) at high levels simultaneously may not be clinically viable for patients with severe lung conditions. Therefore, conventional estimands like the average dose-response function (ADRF), which assumes a counterfactual scenario where all participants receive a specific treatment combination, may be scientifically meaningless due to the impossibility of certain treatment combinations, leading to positivity violations.

Another substantial challenge for vector-valued treatments is the interpretability of the causal estimand. With treatments represented by vectors of continuous variables, the estimated ADRF becomes a high-dimensional surface, complicating intuitive understanding and interpretation. Hence, it is desirable to simplify the complex vector-valued treatment scenario into a more interpretable causal estimand that remains scientifically meaningful. 
Effective estimation of causal effects in the presence of vector-valued treatments is also crucial. Existing inverse probability weighting (IPW) estimators generally require estimating either the generalized propensity score (GPS), which is the conditional density of treatments given covariates, or their ratios. However, this estimation problem becomes particularly challenging when treatments are vector-valued.

\subsection{Our contribution}

In this manuscript, we propose to use modified vector-value treatment policies (MVTP), an extension of modified treatment policies \citep{haneuse2013estimation} for scalar treatments, to handle vector-valued treatments. Beyond this, a key contribution of this manuscript is that we demonstrate the connection between the proposed causal estimand of MVTPs for vector-valued treatments to the average treatment effect on the treated (ATT) for binary treatments, where the binary treatment is an indicator of whether one's treatment has been modified in an expanded population. We prove that both the causal estimands and their required assumptions are equivalent between the causal effects of MVTP and the ATT. This connection enables us to transform the complex, vector-valued treatment scenario into a binary treatment problem, allowing us to directly leverage the vast array of techniques developed in that setting. Following the error decomposition in \citet{jiang2023enhancing} for the finite sample weighted estimator of the causal effect, we propose to use energy balancing weights to estimate the causal estimand of MVTP, as energy balancing weights have been shown to have highly competitive finite sample performance in a variety of settings. Our proposed framework and covariate balancing approach: (1) imposes a less stringent positivity requirement and leverages an energy-distance-based diagnostic tool for empirically assessing positivity and whether the MVTP magnitude is such that measured confounding can be fully controlled; (2) explicitly defines the target population under the MVTP, allowing our energy balancing weights to directly balance covariate distributions without imposing parametric assumptions; and (3) enable a straightforward implementation of the existing sensitivity analysis methods designed for binary treatments.  

Based on the proposed causal MVTP framework, we investigate key scientific questions regarding mechanical ventilation and provide data-driven insights using the MIMIC-III dataset. Specifically, by comparing the causal effect of two different manipulations of MP, we argue that the MP is not the ``final word'' on VILI \citep{vasques2018mechanical} and further provide evidence in support of the medical hypothesis that ``not all the mechanical power are equally hazardous to patients'' \citep{marini2023practical}. In the second case study, we provide a more robust causal analysis of driving pressure, another key ventilation parameter, and mortality.

%%%%   ------------------------------------
%%%%     Input setup section
%%%%   ------------------------------------
\section{The MIMIC-III dataset}
In this paper, we use the MIMIC-III (Medical Information Mart for Intensive Care) dataset \citet{johnson2016mimic} to explore scientifically active problems related to mechanical ventilation. The MIMIC-III dataset is a comprehensive, freely accessible database containing deidentified clinical data of patients admitted to the intensive care units (ICUs) at Beth Israel Deaconess Medical Center between 2001 and 2012. The dataset includes demographics, physiological measurements, laboratory results, medications, and procedures. 
Specifically, for mechanical ventilation research, the dataset provides comprehensive information on ventilator settings, respiratory parameters, and patient outcomes. This enables us to investigate how variations in ventilator management impact clinical outcomes. Notably, the data were collected during a period when there was less consensus on optimal ventilator settings, thereby broadening the treatment domain and mitigating positivity issues in causal effect estimation \citep{del2017mechanical}.  

We follow the approach for cohort construction and data processing outlined in \citet{serpa2018mechanical} to transform the raw MIMIC-III dataset into the structure that is suitable for our data analysis. Among the  over 53,000 ICU admissions in the dataset, the following two inclusion criteria were used: (1) age of 16 years or more; (2) receiving invasive ventilation for at least 48 consecutive hours (in order to control the baseline characteristics), and (3) patients who received ventilation through a tracheostomy cannula were excluded. After applying these criteria, a total of 5011 participants remained for our data analysis. 
  
We use in-hospital mortality as the outcome variable in our analysis. For each patient, we partition the data into two time periods: the first 24 hours and the second 24 hours. The data from the first 24 hours serve as baseline characteristics, which are balanced using our energy balancing weights introduced in Section \ref{sec:3}. A total of 97 baseline covariates are adjusted for to account for potential confounding. The mechanical ventilation data from the second 24 hours are treated as the exposure variables. A detailed definition of the treatment vector for each analysis is provided in Section \ref{sec:5}.

\section{Causal inference for vector-valued treatments}\label{sec:3}

In this section we develop and extend a suite of tools to understand the causal effects of a vector of continuous-valued mechanical ventilator settings on outcomes. We start by introducing modified treatment policies in this context. To then enable the use of a wide variety of techniques for binary treatments, we show the connection between our context and that of binary treatments. This allows us to use flexible tools for binary treatments to control for confounding in complex EHR settings with many confounders. Because unmeasured confounding is possible in the context of mechanical ventilation, sensitivity analysis to this assumption is essential. We then show how the connection between binary treatments and MTPs can allow us to use sensitivity analysis methods developed in the former setting. 

\subsection{Modified vector-value treatment policies (MVTPs)}\label{sec:3.modified}

We extend the modified treatment policy (MTP) framework originally developed for a single continuous treatment \citep{haneuse2013estimation} to accommodate vector-valued treatments. Unlike the commonly used average dose-response function (ADRF), which imagines a counterfactual world where all participants receive the same fixed treatment value $\A=\a$, modified treatment policies focus on a counterfactual scenario where each participant receives a treatment adjusted from their actual (``original") observed treatment value. Specifically, for every participant, this imagined modified treatment is defined as a deterministic function of their original treatment $\boldsymbol{A} = \boldsymbol{a}$ and baseline covariates $\boldsymbol{X} = \boldsymbol{x}$. This particular analyst-specified function is called the treatment policy and is denoted by $(\mathcal{X}, \mathcal{A}) \mapsto \mathcal{A}: \q(\x,\a)$, where $(\mathcal{X},\mathcal{A})$ is the joint support of random variables $(\X, \A)$, and the bold symbol of $q$ indicates that it returns a vector instead of a scalar.

By specifying different treatment policies $\q$, our modified vector-valued treatment policy (MVTP) framework offers substantial flexibility for exploring a wide range of causal questions. For example, suppose we aim to investigate whether ventilator settings characterized by high respiratory rate ($\textnormal{RR}$) and low tidal volume ($\textnormal{TV}$) settings may harm ICU patients, while keeping minute ventilation (defined as $\textnormal{RR}\times \textnormal{TV}$) fixed. In this scenario, we define our vector treatment as $(\textnormal{RR}, \textnormal{TV})$ and choose the corresponding treatment policy as $\q\left(\boldsymbol{X}, (\textnormal{RR}, \textnormal{TV})\right) = (\tau\times \textnormal{RR}, 1/\tau\times \textnormal{TV})$, where the coefficient $\tau$ can be varied to reflect different magnitudes of treatment modification. Although such an MTP/MVTP cannot be exactly implemented in clinical practice, previous studies have demonstrated its scientific value for addressing critical causal questions \citep{jiang2023enhancing,haneuse2013estimation}.

We adopt the potential outcome framework in which we assume that for each subject $i$, there exists a potential outcome $Y^{\a}_i$ defined as the outcome that subject $i$ would have if subject $i$ is intervened on to take the treatment at value $\A = \a$. The estimand of interest for MVTPs is the mean potential outcome $\mu^q\equiv \mathbb{E}[Y^{\q(\X,\A)}]$ under the policy $\q(\x,\a)$, which is then contrasted with the average \textit{observed} outcome $\mathbb{E}[Y]$ to understand whether the policy improves or harms outcomes on average. Identifiability of the causal estimand $\mu^q$ relies on the following assumptions:

\begin{itemize}\label{the:2.consistency}
    \item (A0 Consistency): For any subject $i$ in the population, if $\A_i=\a$ then $Y_i=Y^\a_i$.
     \item (A1 Positivity): If $(\boldsymbol{x},\a)\in (\mathcal{X},\mathcal{A})$ then $(\boldsymbol{x},\q(\boldsymbol{x},\a))\in (\mathcal{X},\mathcal{A})$, where $(\mathcal{X},\mathcal{A})$ is the joint support of random variables $(\X, \A)$.\label{the:2.positivity}
    \item (A2 Conditional exchangeability of related populations) For each $(\boldsymbol{x},\a)\in (\mathcal{X},\mathcal{A})$, let $\a'=\q(\boldsymbol{x},\a)$, then $Y^{\a'}|\X=\boldsymbol{x},\A=\a$ and $Y^{\a'}|\X=\boldsymbol{x},\A=\a'$ have the same distribution.\label{the:2.exchangeability}
\end{itemize}

Assumptions \textbf{A0}–\textbf{A2} extend those introduced by \citep{haneuse2013estimation} to the context of vector-valued treatments. Assumption \textbf{A0} is the standard consistency assumption commonly adopted in causal inference frameworks. Assumption A1, known as the positivity assumption, requires that for each individual in the population, there exist other individuals sharing identical covariate profiles who have actually received the modified treatment. Notably, this positivity assumption is considerably weaker than the one required for estimating the ADRF, where positivity must hold for arbitrary treatment values across the entire population. The conditional exchangeability assumption \textbf{A2} is the MVTP version of the no unmeasured confounding assumption which is much weaker compared with the general no unmeasured confounding assumption for estimating the causal ADRF. We will elaborate more on assumption \textbf{A2} in Section \ref{sec:sensitivity}, where we discuss sensitivity analyses for MVTP.

In order to have a closed-form derivation of the causal estimand $\mu^q$, we further require the treatment policy $q(\x,\a)$ to be block-wise smooth invertible, which is a generalization of the piece-wise smooth invertibility requirement in \citep{haneuse2013estimation}. 
Recall that $\mathcal{A}\subseteq \mathbb{R}^p$ is the support of the vector-value treatment $\A$, then we require
\begin{itemize}
    \item (\textbf{C1 Block-wise smooth invertibility}): For each $\boldsymbol{x} \in \mathcal{X}$, there exists a partition $\{I_{j,\boldsymbol{x}}\}_{j=1}^{J(\boldsymbol{x})} $ of $\mathcal{A}_{\boldsymbol{x}}$ where $\q(\boldsymbol{x},\cdot)$ is smooth and invertible within the partition. Specifically, let $\q_j(\boldsymbol{x},\cdot)$ denote the function $\q(\boldsymbol{x},\cdot)$ on the $I_{j,\boldsymbol{x}}$ part. Then $\q_j(\boldsymbol{x},\cdot)$ has a differentiable inverse function $\h_j(\boldsymbol{x},\cdot)$ on the interior of $I_{j,\boldsymbol{x}}$ (in terms of all partial derivatives), such that $\h_j(\boldsymbol{x},\q_j(\boldsymbol{x},\a))=\a$ for all $\boldsymbol{a}\in I_{j,\boldsymbol{x}}$.\label{the:2.piecewise}
\end{itemize}

Condition \textbf{C1} requires the multivariate policy function $\q(\x,\a)$ to be invertible and smooth with respect to $\a$ within each partition of the support space $\mathcal{A}$. For example, the simplest function that satisfies this condition is the linear transformation function 
$\q(\x,\a) = \boldsymbol{C_x} \a$ where for any given $\x$, $\boldsymbol{C_x}$ be an invertible square matrix transforming the observed treatments to the shifted treatments. The transformation matrix $\boldsymbol{C_x}$ is required to be invertible so that, for any given value of $\x$, $\q(\x,\a)$ have its inverse function $\h(\x,\a)=\boldsymbol{C_x}^{-1}\a$.

Under assumptions \textbf{A0}-\textbf{A2} and condition C1, we can define $\mathbb{E}(Y^{\q(\boldsymbol{x}, \a)}|\X=\boldsymbol{x}, \A=\a)$ to be the expected potential outcome averaging over all subjects who have $\X=\boldsymbol{x}, \A=\a$. The mean potential outcome of the MVTP $\q(\boldsymbol{x},\a)$ is then the mean of the conditional average potential outcome $\mathbb{E}(Y^{\q(\boldsymbol{x}, \a)}|\X=\boldsymbol{x}, \A=\a)$ over the entire population, that is, 
\begin{equation}
    \mu^{q}\equiv\int_{(\mathcal{X},\mathcal{A})} \mathbb{E}[Y^{\q(\boldsymbol{x},\a)}|\X=\boldsymbol{x},\A=\a]dF_{\X,A}(\boldsymbol{x},\a)
\end{equation}
where $F_{\X,\A}(\boldsymbol{x},\a)$ is the distribution function of random variables $(\X,\A)$.
We have the following proposition for the identification result. 
\begin{prop}
    Under assumptions \textnormal{\textbf{A0}}-\textnormal{\textbf{A2}} and condition C1,
    \begin{equation}\label{eq:identification}
    \mu^{q}=\int_{\mathcal{X}} \sum_{j=1}^{J(\x)} \int_{h_j(\x,\a)\in I_{j,\x}} \mathbb{E}(Y|\X=\boldsymbol{x},\A=\a)dF_{\X,\A}(\boldsymbol{x},\h_j(\boldsymbol{x},\a)).
\end{equation}
\end{prop}
A proof is provided in the Supplementary Material.

\subsection{Connection between MVTPs and binary treatments}\label{sec:3.connection}
One distinct advantage of the MVTP framework is that it transforms the complex causal problem for vector-valued treatments into a simple dichotomized problem of whether one adopts the policy or not. This connection is helpful, as provides a formal bridge to methods for binary treatments, allowing their use for complex vector-valued continuous treatments. In this section, we present this connection between the causal effect of an MVTP and the causal effect of binary treatment. By clearly defining the target counterfactual population under an MVTP and providing a clear mathematical expression of it, we show that the causal effect $E[Y^{\q}]-E[Y]$ of an MVTP can be viewed as the causal average treatment effect on the treated (ATT) under a hypothetical binary treatment of whether one adopts the MVTP or not in an expanded population defined later. Although this feature has also been utilized by other authors for the estimation of MTP effects \citep{diaz2021nonparametric}, we believe that we are the first to explicitly articulate and formalize the direct connection between the MVTP effect and the ATT. This explicit connection is useful, as it allows us to straightforwardly borrow techniques for binary treatments, such as sensitivity analyses to unmeasured confounding, for use with vector-valued continuous treatments.

Consider the result \eqref{eq:identification} which represents the causal estimand $\mu^q$ with the integral of the outcome function over some transformed population $F_{\X,\A}(\boldsymbol{x},\h_j(\boldsymbol{x},\a))$ within the partition $I_{j,x}$. With a slight abuse of the notation, we define the indicator function $I_{j,x}(\a)$ as $I_{j,x}(\a)=1$ if and only if $\h_j(\boldsymbol{x},\a)\in I_{j,x}$. Then, the causal estimand $\mu^q$ in \eqref{eq:identification} can be re-written as 
\begin{equation}\label{eq:mu_q}
    \mu^{q}=\int_{(\mathcal{X}, \mathcal{A})} \mu(\boldsymbol{x},\a) dF^{\q}_{\X,\A}(\boldsymbol{x},\a),
\end{equation} 
where $\mu(\boldsymbol{x},\a) \equiv \mathbb{E}(Y|\X=\x,\A=\a)$ is the mean outcome and
\begin{equation}
    F^{\q}_{\X,\A}(\boldsymbol{x},\a)\equiv\sum_{j=1}^{J(\x)} I_{j,x}(\h_j(\x,\a)) F_{\X,\A}(\boldsymbol{x},\h_j(\boldsymbol{x},\a)).
\end{equation}
It can be shown that $F^{\q}_{\X,\A}(\boldsymbol{x},\a)$ is the cumulative distribution function (CDF) of the MVTP transformed variable $(\X,\q(\X,\A))$. Therefore, the causal estimand of MVTP $\mu^q$ can be viewed as the average outcome over the target population $(\X,\q(\X, \A))$ with CDF $F^{\q}_{\X,\A}(\boldsymbol{x},\a)$; and in the context of an IPW style estimator, the weights act to balance the population distribution of $(\X, \A)$ to that of $(\X,\q(\X, \A))$. 

If, instead of viewing $\A$ as treatment variables, we view $\A$ just as additional covariates that have a different distribution between control and target populations, the problem of vector-valued treatments can thus be transformed into a binary treatment problem where $(\X,\A)$ are ``confounders'' with their natural (or ``control'') distribution as that of the observed data and their distribution under the policy as the target (or ``treated'') population. Thus, the causal effect, $\mu^q-\mathbb{E}(Y)$, of an MVTP is just the difference between the average outcomes over the observed population and the target population. To formalize this, consider the augmented population
\begin{equation}\label{eq:aug_population}
    \{\widetilde{\X},\widetilde{\A}, Y, Z\}\equiv\{\X, \A, Y, Z = 0\}\cup\{\X, \q(\X, \A), Y, Z = 1\}
\end{equation}
with $\textnormal{Pr}(Z=1)=0.5$ where $Z=0$ indicates membership in the observed population and $Z=1$ indicates membership in the target. The augmented population \eqref{eq:aug_population} combines the observed population and the target population with an additional binary variable $Z$ indicating the corresponding population. 
If we view $Z$ as a binary treatment variable, then the following lemma shows that the causal estimand of MVTP is equivalent to the causal average treatment effect on the treated (ATT) for the augmented population.

\begin{lemma}\textnormal{(Equivalence of the causal effect between the MVTP and binary treatment)} For the augmented population defined in \eqref{eq:aug_population}, consider $Z$ as the binary treatment and $(\X, \A)$ as the covariates. Then the average treatment effect on the treated (ATT), under the common consistency, positivity, and ignorability assumptions for binary treatment ATT, equals the causal effect of MVTP under assumption \textnormal{\textbf{A0}} -- \textnormal{\textbf{A2}}:
\begin{equation}
    \mathbb{E}(Y(1)-Y(0)|Z=1) = \mathbb{E}(Y)-\mu^q.
\end{equation}
Furthermore, the positivity and ignorability assumptions for binary treatment are equivalent to assumptions \textnormal{\textbf{A1}} and \textnormal{\textbf{A2}} for MVTP. 
\end{lemma}

The equivalence between the ignoability assumption of binary treatment and the assumption \textbf{A2} is in Lemma \ref{the: ignorability}. With this important feature of MVTPs, we demonstrate that: 1) MVTPs have a desirable feature that reduces a complex problem of continuous treatments (or even more complex with vector-valued continuous treatments) to a simple binary treatment problem while still allowing researchers to explore their scientific hypotheses with the original complex treatment structure. 2) The connection between MVTPs and binary treatments allows us to construct an interpretable marginal sensitivity model for MVTPs. Because of this connection, existing approaches can be directly applied to perform sensitivity analyses. Section \ref{sec:sensitivity} gives a more detailed discussion of  sensitivity analysis for MVTPs. 3) Since the treatment variable(s) can be viewed as another ``covariate'' that needs to be balanced to the target population, the extension of MVTPs from single treatment variables to vector-value treatments adds minimum complexity to the estimation procedure. Existing balancing weight methods such as energy balancing can thus be used.

\subsection{Energy balancing tools for MVTP estimation}\label{sec:energybalancingtools}
We propose to estimate the MVTP effect using a set of tools based on the energy distance \citep{szekely2013energy}. Leveraging their connection to binary treatments, it is easy to generalize the energy balancing tools proposed for single-valued continuous treatments \citep{jiang2023enhancing} to the multivariate setting. Specifically, our tools enable (1) nonparametric estimation of optimal balancing weights for MVTP and (2) selection of a feasible MVTP scale to ensure the validity of the positivity assumption and that measured confounding is fully controlled. In the Supplementary Material, we introduce the core ideas of the proposed energy balancing tools for vector-valued treatments. For detailed descriptions of the methods and the corresponding theoretical results, we refer readers to \citet{jiang2023enhancing}. In Section \ref{sec:5}, we evaluate the practical performance of the energy balancing methods through plasmode simulations based on real data.

\subsection{Sensitivity analysis for MVTPs}\label{sec:sensitivity}

The estimation of causal effects from observational data relies on untestable assumptions. Sensitivity analysis, which evaluates the robustness of the causal conclusions to potential violations of assumptions, is thus critical for interpreting the estimated causal effects in real-world applications. When estimating the causal effect of an MVTP, a key untestable assumption is the conditional exchangeability of the relevant populations (Assumption \textbf{A2} in Section \ref{sec:3.modified}). 
In Section \ref{sec:3.connection}, we establish that the causal effect of an MVTP for a continuous treatment is equivalent to the ATT for a binary pseudo-treatment. In this subsection, we demonstrate that the assumption \textbf{A2} is also equivalent to the general ignorability (``no unmeasured confounding") assumption for binary treatments. This equivalence also has a significant implication: it enables us to construct an interpretable marginal sensitivity model assess the assumption \textbf{A2} for MVTPs when the treatment is a single/vector-value of continuous variable(s). Furthermore, we show that this marginal sensitivity model can be readily utilized using existing methods developed for sensitivity analysis of binary treatments.

\subsubsection{A marginal sensitivity model for MVTPs}\label{sec:3.sensitivity}

Recall that conditional exchangeability of the relevant populations assumption requires for each $(\x,\a)\in(\mathcal{X},\mathcal{A})$, $Y^{\a'}|\X=\x,\A=\a$ and $Y^{\a'}|\X=\x,\A=\a'$ have the same distribution where $\a'=q(\x,\a)$ is the modified treatment. This assumption states that $\X$ includes all variables that simultaneously 1)  impact the outcome and 2) have a different distribution in the observed (natural) population from the target (modified) population. In the degenerate case that $\a=q(\x,\a)$ always, the assumption holds by default. 
This gives us an intuition on the following connection between the conditional exchangeability assumption \textbf{A2} and the ignorability assumption for the ATT for binary treatments. Although \textbf{A2} is equivalent to no unmeasured confounders for the pseudo-treatment $\Z$, it is in general weaker than standard no unmeasured confounding because the natural and modified populations are likely to be similar for policies that do not modify the treatment dramatically.

\begin{lemma}\textnormal{(Equivalence between the assumption \textbf{A2} and the ignorability assumption)} \label{the: ignorability}
For the augmented population defined in \eqref{eq:aug_population}, consider $Z$ as the binary treatment and $(\X, \A)$ as the covariates. Then the ignorability assumption for the ATT given $Z$ is equivalent to the conditional exchangeability of related population assumption \textnormal{\textbf{A2}} for MVTPs. 

\end{lemma}

Based on this equivalence between assumption \textbf{A2} and the ignorability assumption for binary treatment, we can construct a marginal sensitivity model for MVTPs. Considering the augmented population \eqref{eq:aug_population}, for the pseudo treatment variable $Z$, we can define the observed ``propensity score" as $\pi(\x,\a)=\textnormal{Pr}(Z=1|\X=\x,\A=\a)$ and the ``true propensity score" as $\pi(\x,\a,y)=\textnormal{Pr}(Z=1|\X=\x,\A=\a,Y(1)=y)$. Then, the marginal sensitivity model for the causal effect of the MVTP can be defined as:

\begin{itemize}\label{the:A3}
    \item (A3 Marginal sensitivity model for MVTP): For $\Lambda\geq1$, the true ``propensity score" for the pseudo treatment $Z$ satisfies
    \begin{equation*}
        \pi^*(\x,\a,y)\in\mathcal{E}(\Lambda)=\{\pi^*(\x,\a,y)\in(0,1):\Lambda^{-1}\leq \textnormal{OR}(\pi(\x,\a),\pi^*(\x,\a,y))\leq\Lambda\}
    \end{equation*}
    where $\textnormal{OR}(\pi(\x,\a),\pi^*(\x,\a,y))=\frac{\pi(\x,\a)/(1-\pi(\x,\a))}{\pi^*(\x,\a,y)/(1-\pi^*(\x,\a,y))}$ is the odds ratio of the twos ``propensity scores".
\end{itemize}

Since $Z=0$ corresponds with the observed population of $(\X, \A)$ and $Z=1$ corresponds with the MVTP-shifted target population $(\X, \q(\X,\A))$, the observed propensity score can be represented as 
\begin{equation}
    \pi(\x,\a)=\textnormal{Pr}(Z=1|\X=\x,\A=\a)=\frac{d F^q_{\X,\A}(\x,\a)}{ dF_{\X,\A}(\x,\a)+dF^q_{\X,\A}(\x,\a)},
\end{equation}
and the true propensity score can be represented as 
\begin{equation}
    \pi^*(\x,\a,y)=\textnormal{Pr}(Z=1|\X=\x,\A=\a,Y(1)=y)=\frac{dF^q_{\X,\A,Y}(\x,\a,y)}{dF_{\X,\A,Y}(\x,\a,y)+dF^q_{\X,\A,Y}(\x,\a,y)}.
\end{equation}
where $F^q_{\X,\A,Y}$ is the distribution for $(\X, \q(\X, \A), Y)$. Based on this, the marginal sensitivity model \textbf{A3} for MVTPs has the following more interpretable expression.

\begin{lemma}\textnormal{(Connection between the marginal sensitivity model and ratio of the weights)} Denote the true data-generating mechanism as $\mathbb{P}(\X,\A,Y,\U)$ where $\U$ is the unobserved confounders such that the conditional exchangeability of related populations assumption \textnormal{\textbf{A2}} holds given $\X$ and $\U$. Then for $\Lambda\geq 1$, the marginal sensitivity model \textnormal{\textbf{A3}} is equivalent to assuming the true data-generating mechanism $\mathbb{P}(\X,\A,Y,\U)$ follows:
\begin{equation}
    \mathbb{P}(\X,\A,Y,\U)\in\mathcal{P}(\Lambda)=\{\mathbb{P}(\X,\A,Y,\U):\Lambda^{-1}\leq  \frac{w}{w^*} \leq\Lambda \}
\end{equation}
where $w$ is the density ratio weights under the observation of $(\X, \A)$ and $w^*$ is the ``true" density ratio weights if we observe $(\X, \A, \U)$.
\end{lemma}

\subsubsection{Solving for the sensitivity parameter}\label{sec:solve.sensi}
Recent work has advanced sensitivity analysis methods for binary treatments with weighting
estimators. \citet{zhao2019sensitivity} proposed the marginal sensitivity model for inverse propensity weighting estimators. They estimate the confidence interval using a percentile bootstrap and a generalized minimax-maximin inequality. \citet{dorn2025doubly,dorn2023sharp}; and \citet{tan2024model} further introduced approaches to estimate sharp confidence intervals under the marginal sensitivity model. \citet{soriano2023interpretable} proposed a generalizable percentile bootstrap-based sensitivity analysis for a broad class of balancing weights and introduced an amplification technique to enhance interpretability. 

The only difference between the marginal sensitivity model \textbf{A3} for MVTPs and the marginal sensitivity model for binary treatments (Definition 1 in \citet{zhao2019sensitivity}) is that the vector-valued treatment variable $\A$ is viewed as merely other observed ``covariates" in the sensitivity model. Therefore, the existing approaches for estimating the confidence interval of the causal effect under the marginal sensitivity model with parameter $\Lambda$ can be directly applied to our MVTP framework. We adopt the doubly-valid/doubly-sharp sensitivity analysis approach proposed by \citep{dorn2025doubly} for solving for the sensitivity parameter, as it provides sharper and more precise intervals. We note that during the writing of this manuscript, we found that \citet{levis2024stochastic} proposed a sensitivity framework for ``generalized policies'' which contains the non-stochastic intervention MTP as a special case.

\section{Simulation experiments}\label{sec:4}
In this section, we study the empirical properties of the proposed energy balancing weights compared with other commonly used estimators for the estimation of MVTPs. The aim of the simulation study is to evaluate the performance of energy balancing estimators under various simulation conditions in comparison with competing approaches, especially in scenarios similar to our mechanical power of ventilation case study. 

\subsection{Data generating mechanisms}

We adopt the plasmode simulation strategy \citep{vaughan2009use,franklin2014plasmode} for our simulation experiment. Plasmode simulation is a data simulation technique that utilizes real-world data to evaluate the performance of statistical methods under realistic conditions. Unlike fully synthetic simulations, which generate data from theoretical distributions, plasmode simulations modify observed data while retaining its inherent structure and correlations. In this experiment, we simulate patient covariates ($\X$) and treatment levels ($\A$) according to the observed MIMIC-III data, which preserves the underlying correlation between the covaraites and treatment variables. The outcome variable ($Y$) is then generated using a known function of $\X$ and $\A$ with a specified level of variation. Given the complexity of the relationship between covariates and treatment in our real-world application, this data-generation approach enables a more realistic evaluation of the performance of different methods in practice. \citet{shaw2025cautionary} showed that plasmode simulations yield non-negligible positivity violations when treatment variables are directly drawn from the data. To mitigate this, we introduced a small amount of additional randomness to the treatment variables (specifically, adding noise with a standard deviation equal to 5\% of the treatment variable’s standard deviation) to mitigate the risk of such positivity violations.

We vary the following settings for the simulation experiment: the sample size ($n = 100, 200, 400, 800$), the number of covariates ($p = 10, 20, 40, 80$) for each individual, and the magnitude of MTP shift ($\tau=0.05, 0.1, 0.15, 0.2$, explained later). For each simulation scenario, denote the subset of our real data as $\{(\X_{(p)},\A)\}$ where $\X_{(p)}$ is the first $p$ variables of the patient characteristics, and $\A$ is the treatment vector with 5 variables. We then randomly draw $n$ observations from the real data $\{(\X_{(p)},\A)\}$ with replacement and add additional noise to the treatment variables, denoted as $\{(\X_{(p)i},\A_i)\}_{i=1}^n$. The outcome variable $Y_i$ is then generated follow a Normal distribution with mean $\mu(\X,\A)$ and variance 1. The mean function $\mu(\X,\A)$ is a complex polynomial of $\X$ and $\A$, see the Supplementary material for the specific expression. The MVTP shift function $q$ is designed to be $q(\X,\A)=(1-\tau)\A$ where $\tau$ controls the magnitude of MVTP shift.

\subsection{Estimands and estimators under evaluation}

Due to the connection between the causal effect of MVTPs and the ATT of binary treatments, it is straightforward to extend other existing approaches for MTP with single level treatment to vector-valued treatments. In the simulation experiment, we compare the following methods: (1) the penalized energy balancing estimator, as introduced in Section \ref{sec:energybalancingtools} with the penalty factors $\lambda$ set to be $1$ for all simulation scenarios,  (2) a kernelized energy balancing approach using the maximum mean discrepancy (MMD) with a Gaussian kernel; see Supplementary Material Section 2.2 of \citet{jiang2023enhancing}, (3) the naive, unadjusted method that assigns equal weights to all subjects, and the classification method proposed by (\cite{diaz2021nonparametric} Section 5.4 with classification models estimated by either (4) a logistic regression or (5) a random forest. Except for the pure weighting approaches for covariate balancing, we also implement augmented estimators for each weighting method using the same outcome model, an ensemble method SuperLearner \citep{van2007super} that incorporates  lasso regularized generalized linear models (\texttt{SL.glmnet}) and random forests.

For each method, the estimand under evaluation is the mean potential outcome under the MVTP, $\mu^{\q}$. We present both the bias and coverage rate for the 95\% confidence interval for each method, with confidence intervals calculated using the non-parametric bootstrap. 

\begin{figure}[ht]
    \centering
     \includegraphics[width=\textwidth]{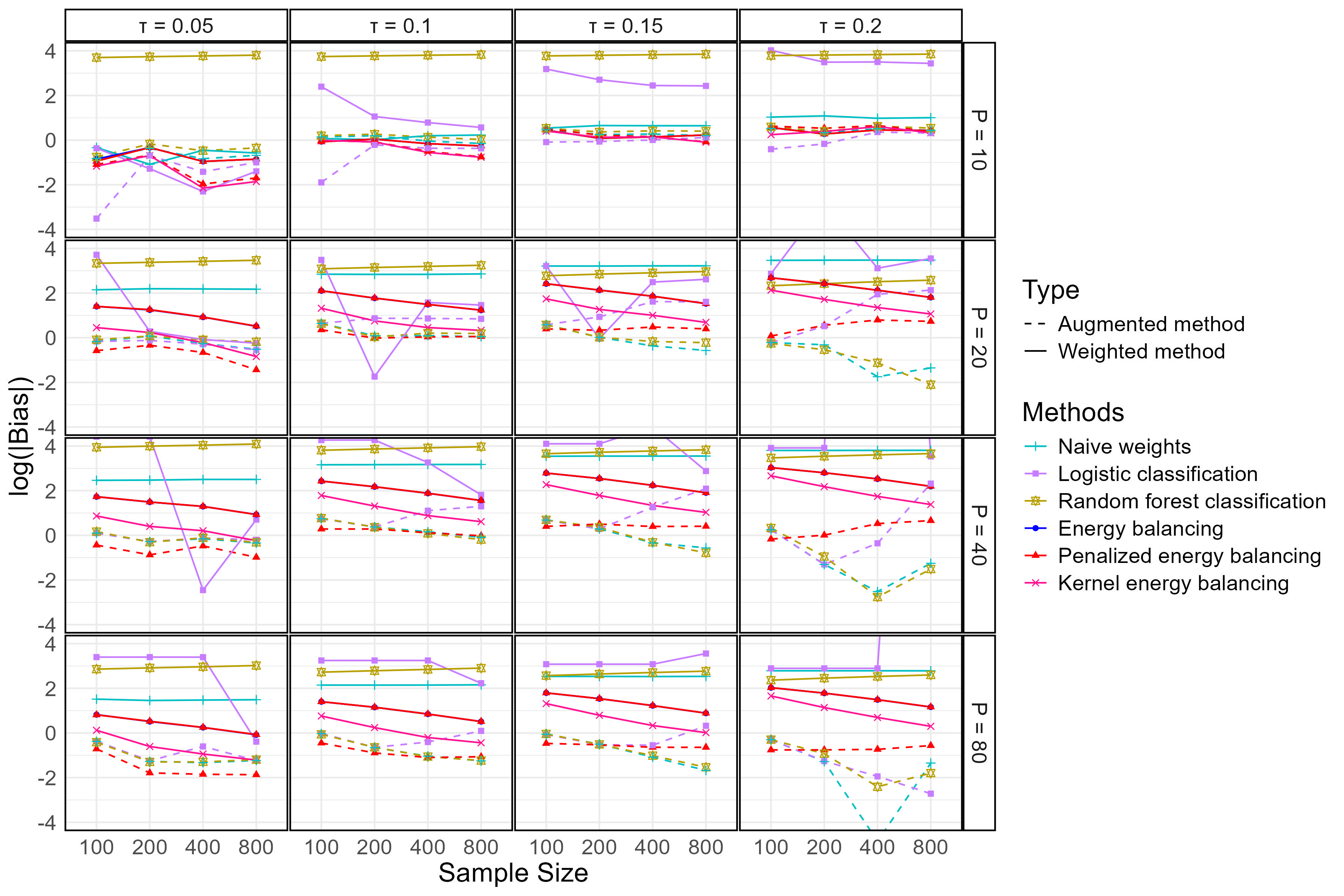}
    \caption{Simulation results in terms of the logarithm of the absolute value of bias across different sample sizes, magnitude of MVTP shift, and the dimensionality of covariates. Balancing methods are displayed in different colors and shapes of points. Weighted estimators are displayed in solid lines and the augmented estimators are in dashed lines.}
    \label{fig:simulation_bias}
\end{figure}

\subsection{Simulation results}

\begin{figure}[ht]
    \centering
     \includegraphics[width=\textwidth]{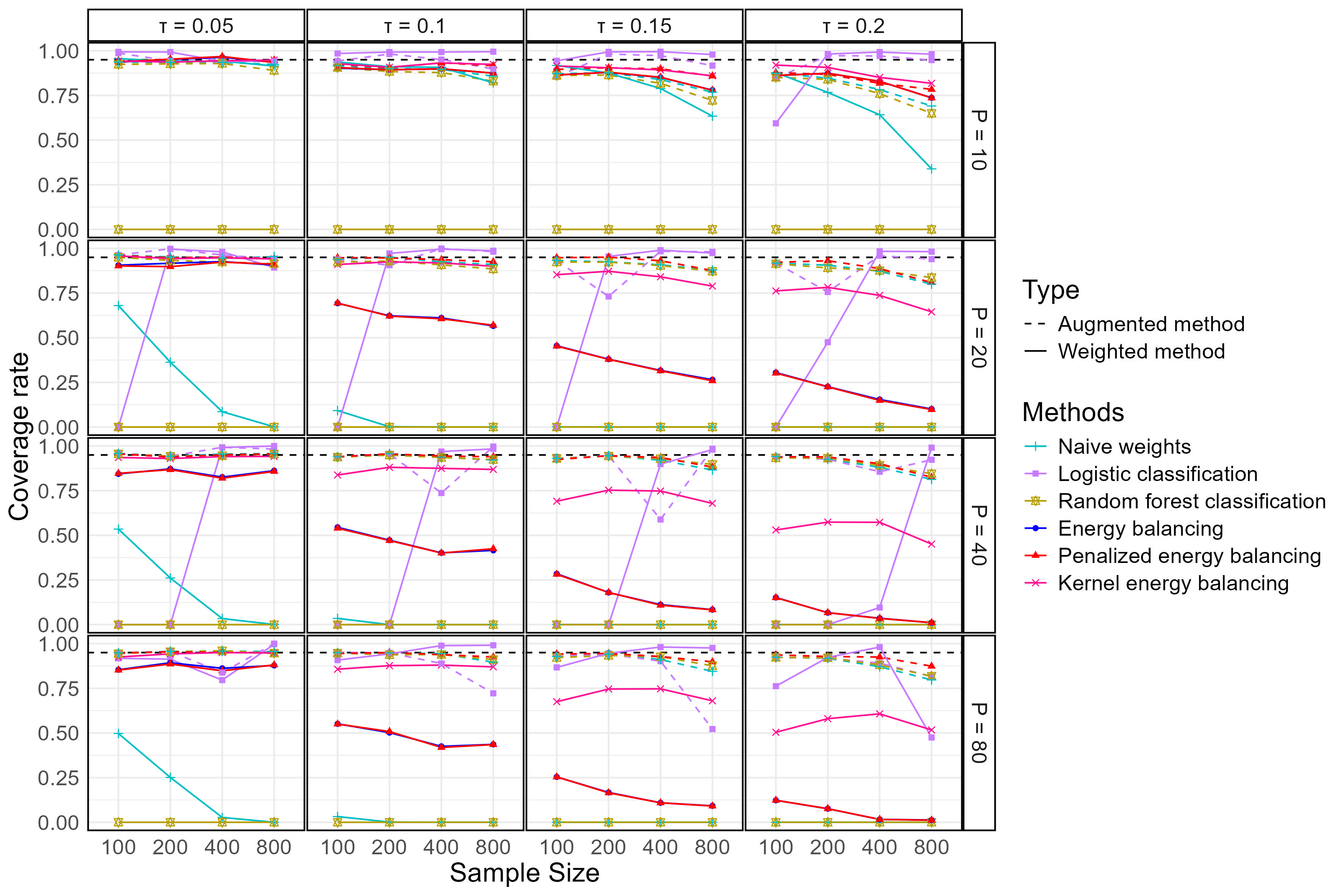}
    \caption{Simulation results in terms of the coverage rate of the 95\% confidence interval across different sample sizes, magnitude of MVTP shift, and the dimensionality of covariates. Balancing methods are displayed in different colors and shapes of points. Weighted estimators are displayed in solid lines and the augmented estimators are in dashed lines.}
    \label{fig:simulation_coverage}
\end{figure}

The simulation results are shown in Figure \ref{fig:simulation_bias} (log absolute bias) and Figure \ref{fig:simulation_coverage} (coverage rate of the 95\% confidence interval). Figure \ref{fig:simulation_bias} illustrates that estimation bias for the weighting method generally increases with the magnitude of the MVTP shift, $\tau$. This trend aligns with expectations, as larger shifts make covariate balancing more challenging and exacerbate positivity issues, particularly when all five treatment variables are shifted simultaneously. Among the pure weighting methods, our energy balancing approaches (penalized energy balancing and kernel energy balancing) consistently perform well across nearly all scenarios. Kernel energy balancing has the best performance in most cases with $p > 10$. The classification method using logistic regression shows low bias in some cases (e.g., $\tau = 0.1, p = 20, n = 200$), but its performance is highly variable, with some scenarios yielding extremely poor results under the same $\tau$ and $p$ but different sample sizes, $n$. All augmented estimators reduce bias compared to their corresponding pure weighting methods, with our augmented energy balancing estimator demonstrating robust performance across scenarios.

For the coverage rate in Figure \ref{fig:simulation_coverage}, all the pure weighting methods have a bad coverage rate with a large shift magnitude $\tau>0.15$ and high dimensionality $p>20$. However, our energy balancing methods (particularly the kernel energy balancing method) still have a better coverage rate compared with other methods. The classification method with logistic regression sometimes has a 100\% coverage due to its extremely large variability under bootstrap. With the help of the outcome regression, the coverage rates for all the augmented methods are generally improved. Our augmented energy balancing method consistently has the best coverage rate in almost all the scenarios.

\section{Exploring ventilator-induced lung injury (VILI)}\label{sec:5}

Using our MVTP framework, we investigate the key scientific questions raised earlier and provide data-driven insights using the MIMIC-III dataset. Specifically, in this section, we focus on: (1) validating that different combinations of the five components contributing to total mechanical power may be differentially hazardous to patients, even if they yield  the same mechanical power; (2) exploring the effect of driving pressure (DP) on patient mortality. As suggested by the simulation experiments, we use the augmented energy balancing estimator for estimating the average potential outcome under MVTP where the weights are generated by the kernel energy balancing method with Gaussian kernel, and the outcome regression functions are estimated by SuperLearner with Lasso (\texttt{SL.glmnet}) and random forest (\texttt{SL.ranger}) models. To assess sensitivity of findings to unmeasured confounding, we solve for the largest value of the sensitivity parameter $\Lambda$ such that the results are still statistically significant. The largest sensitivity parameters are calculated using the doubly-valid/doubly-sharp (DVDS) sensitivity analysis approach as described in Section \ref{sec:solve.sensi} and in the Supplementary Material. 

\subsection{Studying whether components of MP are  equally hazardous}
\begin{figure}[ht]
 \centering
    \includegraphics[width=\textwidth]{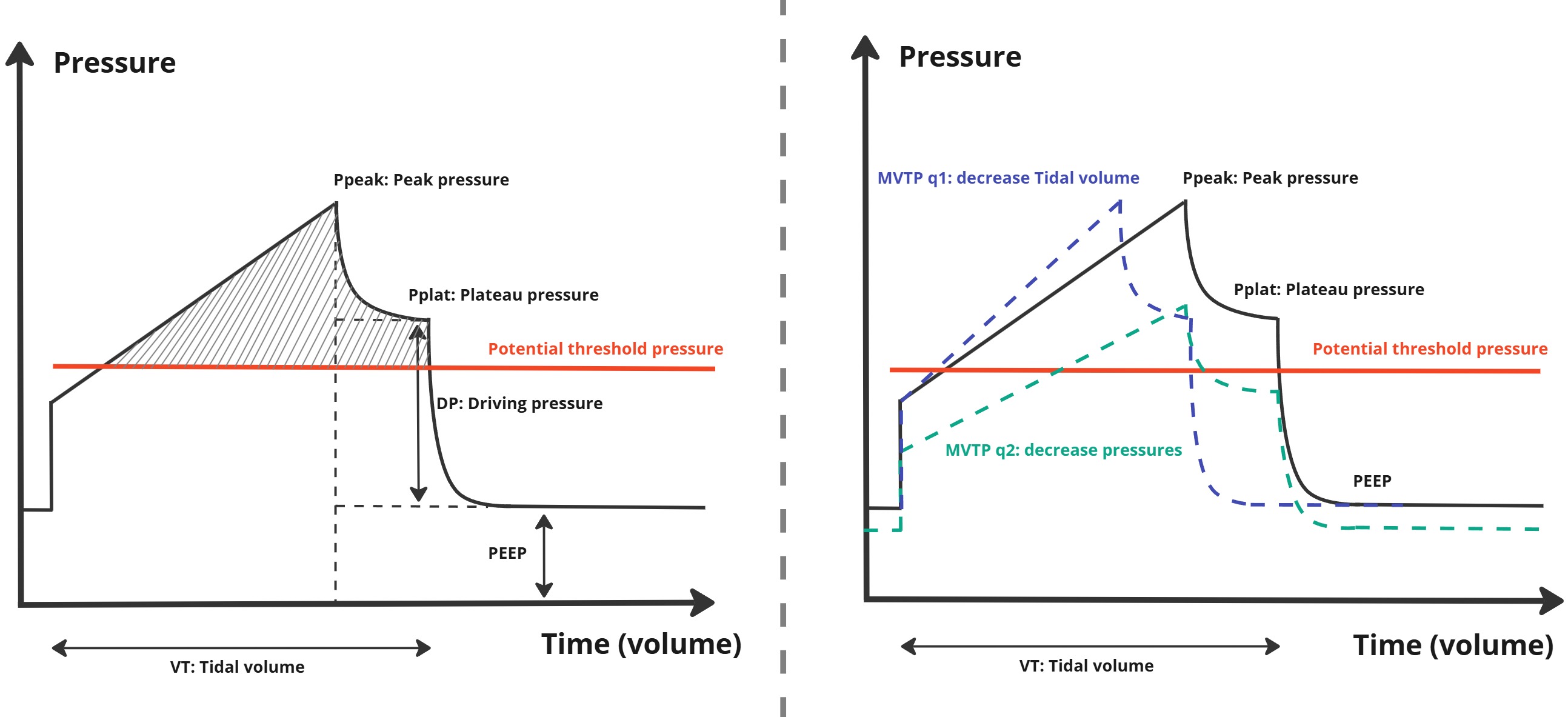}
    \caption{Illustration of airway pressure during a single ventilation cycle. The X-axis represents time, which correlates with the volume of air entering the patient's lungs. The Y-axis represents lung pressure. The blue dashed line indicates the subject-specific pressure threshold.}
    \label{fig:MPintro}
\end{figure}

Although MP has been observed to be positively associated with mortality in critically ill patients \citep{neto2018mechanical}, its underlying mechanism remains unclear. The left panel of Figure \ref{fig:MPintro} illustrates the airway pressure waveform during a ventilation cycle. The X-axis represents time, which also correlates with the volume of air delivered to the patient’s lungs, as mechanical ventilation typically pushes air at a relatively constant rate. The Y-axis represents airway pressure, capturing fluctuations throughout the respiratory cycle. The figure also highlights $P_{\textnormal{peak}}$, $P_{\textnormal{plateau}}$, and PEEP that contribute to the overall mechanical power and influences the mechanical stress and strain exerted on the lungs during ventilation. For a fixed $\textnormal{RR}$, MP can be viewed as the area under this airway pressure curve. 

It has been hypothesized that different combinations of these five components, even when resulting in the same total MP, may have varying effects on patient outcomes \citep{marini2019optimize, marini2020component}. In particular, \citet{marini2020component} suggested that each patient has a subject-specific pressure threshold (represented by the red line in Figure \ref{fig:MPintro}). They further hypothesized that, instead of the total power (mechanical power), what really causes lung injury is the power that exceeds the pressure threshold (the shaded area in the left panel of Figure \ref{fig:MPintro}). While this hypothesis is medically plausible, its statistical validation using observational data is challenging due to the inherent complexity. In this subsection, we aim to shed light on this hypothesis using our proposed modified vector-valued treatment policy (MVTP) framework and the corresponding energy balancing methods.

We define the ``treatment'' vector as the five components of mechanical power: 
\begin{equation}
    \A = (\textnormal{RR}, V_T, P_{\textnormal{peak}}, P_{\textnormal{plateau}}, \textnormal{PEEP}). 
\end{equation}
It is important to note that not all five components are directly modifiable in clinical practice. Clinicians can adjust $\textnormal{RR}$, $V_T$, and PEEP. However, the pressure variables, $P_{\textnormal{peak}}$ and $P_{\textnormal{plateau}}$, are inherently linked to these modifiable variables, meaning that changing one will influence the others. Despite this, the MVTP framework is useful, as it allows comparisons between observed and counterfactual treatment scenarios in which only the treatment vector $\mathbf{A}$ differs.

For each $0<\tau\leq1$, we consider two MVTPs $q_1(\X, \A)$ and $q_2(\X, \A)$ that aim to provide insights into the causes of VILI, both of which modify total mechanical power to $\tau$ times its original value, but through different ventilator components:
\begin{equation}
    q_1(\X, \A) = (\textnormal{RR}, \tau\times V_T, P_{\textnormal{peak}}, P_{\textnormal{plateau}}, \textnormal{PEEP})
\end{equation}
and the MVTP $q_2$ is:
\begin{equation}
    q_2(\X, \A) = (\textnormal{RR}, V_T, \tau\times P_{\textnormal{peak}}, \tau\times P_{\textnormal{plateau}}, \tau\times \textnormal{PEEP})
\end{equation}
The MVTP $q_1$ modifies the tidal volume $V_T$ by scaling it to $\tau$ times its original value, while the MVTP $q_2$ modifies the three airway pressure variables -- $P_{\textnormal{peak}}$, $P_{\textnormal{plateau}}$, and $\textnormal{PEEP}$ to $\tau$ -- by the same factor $\tau$. An illustration of MVTPs $q_1$ and $q_2$ is provided in the right panel of Figure \ref{fig:MPintro}. The blue dashed line represents the potential airway pressure curve under the MVTP $q_1$ where we decrease the tidal volume $V_T$ but fix the airway pressure variables. The green dashed line represents the potential airway pressure curve under $q_2$ where we decrease the airway pressure variables but fix the tidal volume $V_T$. It is apparent from Figure \ref{fig:MPintro} that MVTP $q_2$ has a much smaller ``hazardous power'' (that is, power that exceeds the threshold pressure) than the MVTP $q_1$. Following the hypothesis proposed by \citet{marini2020component}, although both MVTPs modify mechanical power by the same magnitude, their effects on the lungs should be different. Since lung damage is hypothesized to be primarily associated with power exceeding a threshold pressure, the MVTP that directly modifies the airway pressures ($P_{\textnormal{peak}}$, $P_{\textnormal{plateau}}$, and $\textnormal{PEEP}$) is expected to have a greater impact on the ``hazardous power'' and, consequently, on patient mortality compared one that modifies $V_T$.

\begin{figure}[h]
 \centering
    \includegraphics[width=\textwidth]{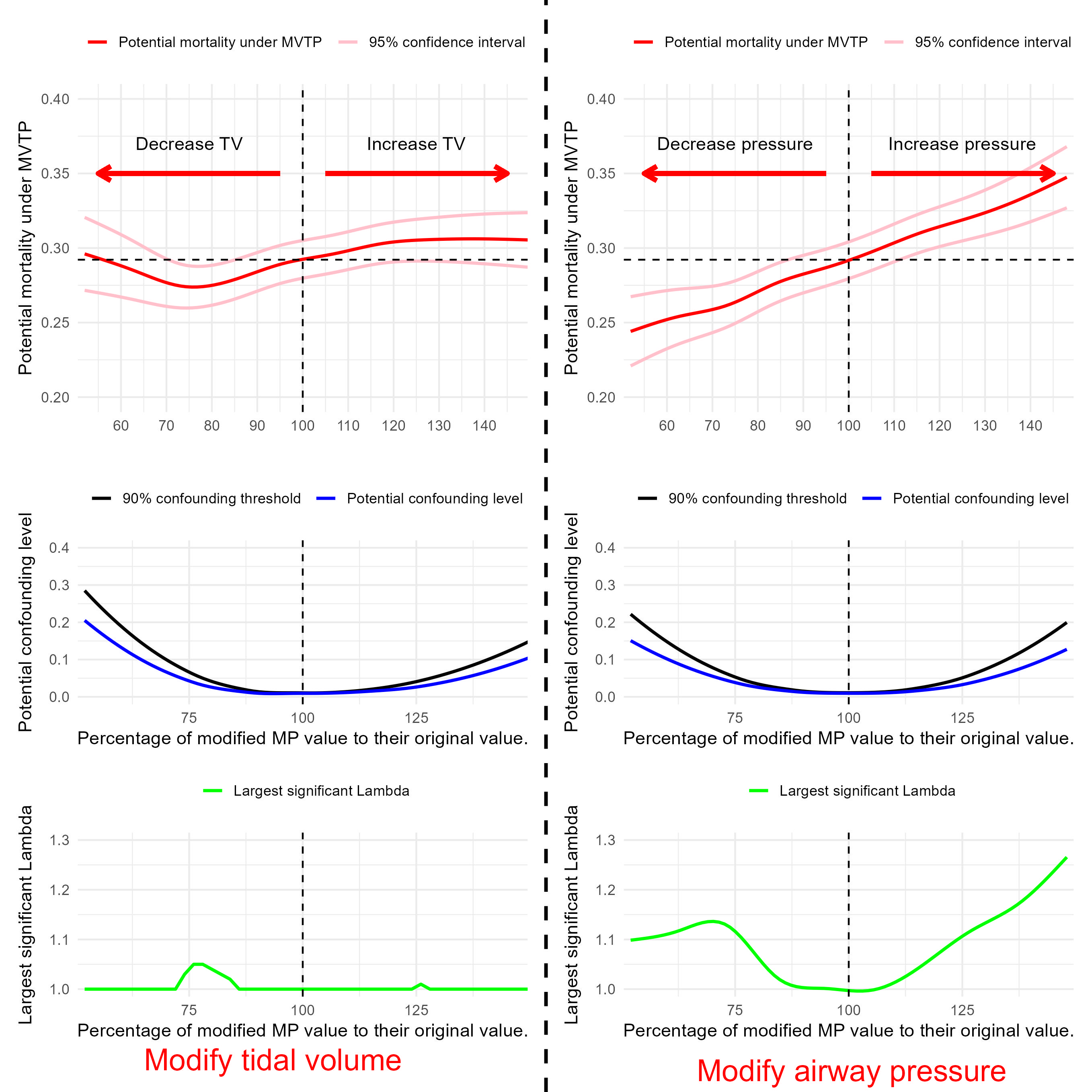}
    \caption{Exploring the potential outcome under different magnitudes of shifts of tidal volume and peak pressure. The X-axis represents the $100\times\tau$ value of the mechanical power of the modified treatments to the original mechanical power. The MVTP on the left panel ($q_1$) modifies tidal volume $V_T$, while the MVTP on the right panel modifies the airway pressure variables (peak pressure, plateau pressure, and PEEP). The red curve is the estimated mortality with the corresponding shifts. Two pink curves represent the 95\% confidence interval of the mortality. The black curve on the bottom represents the threshold for the permutation test of confounding control and the blue curve is the remaining confounding level after covariate balancing. As suggested by the middle plots, measured confounding is adequately controlled for all $\tau$. The green curve presents the largest $\Lambda$ value for which the causal effect remains significant under the marginal sensitivity model.}
    \label{fig:casestudy1}
\end{figure}

Figure \ref{fig:casestudy1} presents the estimated causal effects for the MVTP $q_1$ and $q_2$ as $\tau$ ranges from $0.5$ to $1.5$. The left panel shows the results for MVTP $q_1$, where only $V_T$ is modified, while the right panel displays the results for $q_2$, in which all three airway pressure variables are simultaneously modified. In the middle panel, the black curve on the bottom represents the threshold for the permutation test of confounding control and the blue curve is the observed test statistic after covariate balancing. Since the blue curve never exceeds the black, confounding is adequately controlled for all $\tau$. The findings indicate that decreasing the three pressure variables ($\tau < 1$) can significantly reduce the potential in-hospital mortality rate (as shown by the red curve). On the other hand, increasing the pressure variables (for $\tau>1$) significantly increases mortality, validating the hazardous nature of the airway pressure variables. In contrast, modifying $V_T$ appears to have negligible effect on mortality, as the estimated average potential mortality shows minimal change, and the confidence intervals suggest that increasing $V_T$ does not have a significant impact on mortality. 

Using the marginal sensitivity framework described in Section \ref{sec:3.sensitivity} and the DVDS sensitivity analysis approach, we estimate the largest $\Lambda$ value for which the MVTP effect remains significant; a larger $\Lambda$ indicates greater robustness against potential unmeasured confounders. The sensitivity analyses results further support our previous observations, where modifying the pressure variables has a much larger $\Lambda$ value compared with the modification of tidal volume, indicating moderate robustness of the findings to unmeasured confounding. Overall, these results support the hypothesis that it is not the total mechanical power, but rather the power exceeding the pressure threshold, that poses the greatest risk to lung health.

\subsection{The role of driving pressure}

Our next investigation examines the lung protection effect of lowering the driving pressure (DP) of a ventilator. For context, in the clinical setting, the provider prescribes a ventilation mode that targets tidal volume and regulates either gas flow or airway pressure. With advancing knowledge, the approach to lung protective ventilation progressed from restricting tidal volume, to minimizing alveolar collapse, to avoiding recurrent tidal recruitment. Following publication of an influential analysis of high-quality clinical trials data linked to mortality \citep{amato2015driving}, attention currently centers on restricting the excursion of airway pressure--the driving pressure--needed to accomplish safe tidal gas delivery. DP is defined as the difference between $P_{\textnormal{plateau}}$ and $\textnormal{PEEP}$--two adjustable pressures--and has been previously associated with patient mortality \citep{amato2015driving}. Yet, because the stimulus for VILI is believed to related to excessive strain (`stretch') at the parenchymal level, this driving pressure excursion itself cannot precisely index unsafe ventilation, as the `stretch' that occurs in response to DP depends upon the local elastance and vulnerability of the respiratory system \citep{goligher2021effect}. Thus, for ventilated patients with no lung injury, a high tidal volume or even a high DP applied to a lung with a stiff surrounding chest wall may not stretch alveoli enough to
cause VILI. However, the same tidal volume or DP may prove unsafe for those with acute respiratory distress syndrome (ARDS), a life-threatening lung condition leading to severe respiratory distress, and a normal chest wall who are more likely to have a small `baby lung' whose fewer open alveoli are more susceptible to harmful stretch \citep{marini2020time}. 

In this section we aim to investigate the link between DP and mortality considering the above. Yet, the volume excursion that occurs in response to the DP must also be considered, and that excursion requires a mechanical energy input, which is a pressure-volume product. Importantly, driving pressure is also inherently linked to disease severity through the relationship $\textnormal{DP} = V_T / \textnormal{C}{rs}$, where $V_T$ represents tidal volume, and $\textnormal{C}{rs}$ denotes respiratory-system compliance, the inverse of elastance. For the same chest wall, a higher $\textnormal{C}_{rs}$ indicates better lung compliance, more functional lung units, and consequently, less severe illness. Therefore, the observed association between lower driving pressure and reduced mortality may simply reflect differences in disease severity rather than a direct causal effect. This distinction is clinically meaningful: if reducing driving pressure itself improves outcomes, targeted interventions based on DP can be designed; if not, simply lowering DP without addressing underlying lung compliance may offer limited benefit. 
 
We aim to disentangle the influence of $\textnormal{C}{rs}$ and assess the causal impact of modifying driving pressure on patient mortality. To do so, we first define the treatment vector as $\A = (\textnormal{RR},V_T, \textnormal{DP})$.
Then for each $\tau$, we consider the MVTP $q_3(\X,\A)$ such that:
\begin{equation}
    q_3(\X,\A) = \Big(\frac{1}{\tau}\times\textnormal{RR}, \tau \times V_T, \tau \times\textnormal{DP}\Big).
\end{equation}
Because $q_3(\X,\A)$ proportionally modifies both tidal volume and driving pressure by $\tau$ and simultaneously decreases the respiration rate by $1/\tau$, it inherently: (1) preserves $\textnormal{C}{rs} = V_T / \textnormal{DP}$; and (2) preserves the minute ventilation defined as the product of $\textnormal{RR}$ and $V_T$, both of which measure the severity of patient illness. Therefore, the MVTP $q_3(\X,\A)$ ensures that patient severity remains unchanged before and after the modification. By comparing the average potential outcomes under MVTP $q_3(\X,\A)$ to the observed outcomes, we can investigate whether reducing driving pressure has a direct causal effect on improving patient mortality. In addition to estimating the causal effect of the MVTP $q_3(\X,\A)$ over the whole ICU patients in MIMIC-III dataset, we also estimate the causal effect to a more restricted population who have been diagnosed with ARDS with the hypothesis that modifications to DP for the general ventilated population may have limited impact, but may have a harmful effect among vulnerable patients with ARDS.

\begin{figure}[h]
 \centering
    \includegraphics[width=\textwidth]{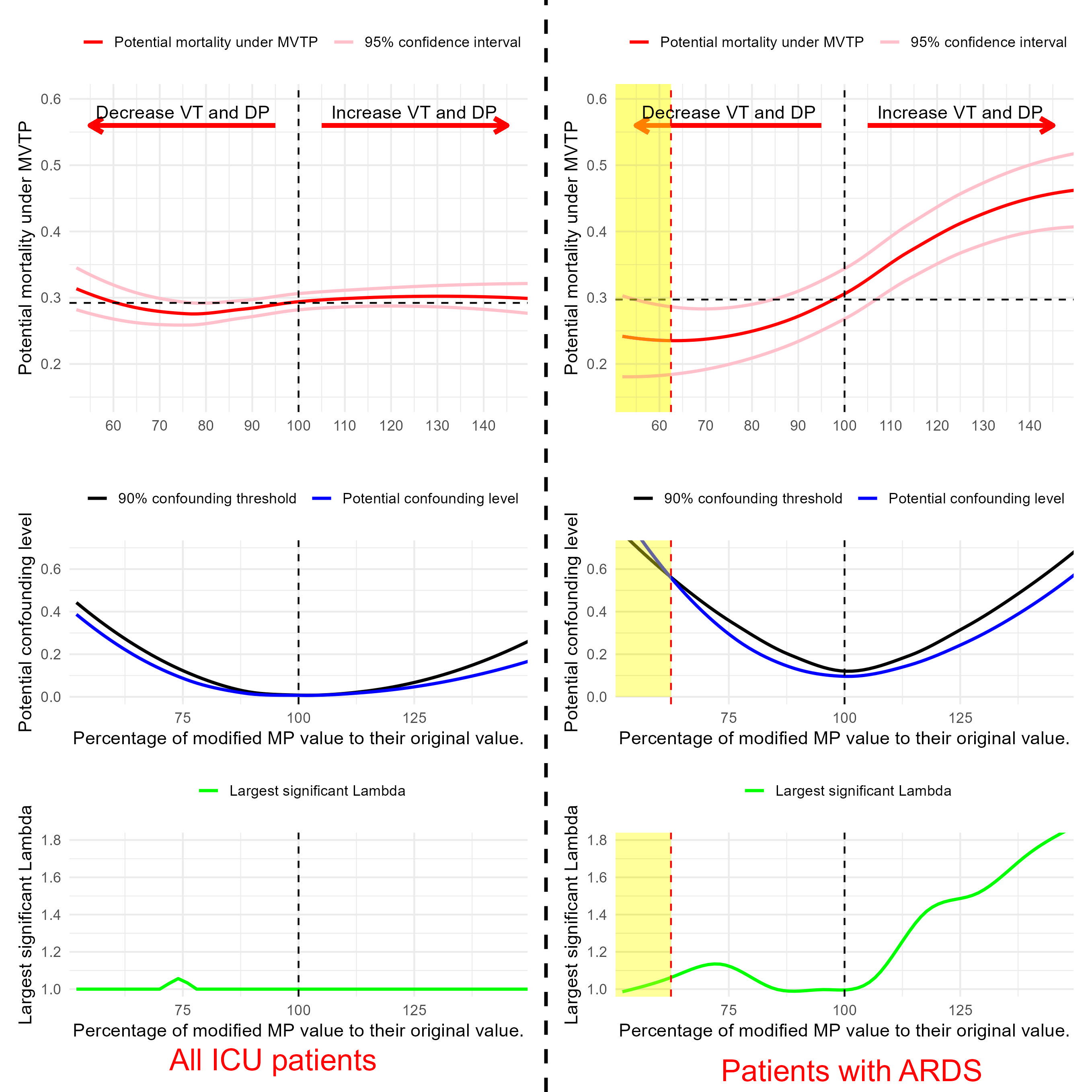}
    \caption{Exploring the potential outcome under different magnitudes of shifts of tidal volume (VT), driving pressure (DP), and respiratory rate (RR). The X-axis represents the $100\times\tau$ value of the modified tidal volume and driving pressure to their original value. The area to the left of the dashed vertical line decreases VT and DP while increasing RR; the right area increases VT and DP while decreasing RR. The red curve is the estimated mortality with the corresponding shifts. Two pink curves represent the upper and lower bounds of the 95\% confidence interval of the mortality. The black curve on the bottom represents the threshold for the permutation test of confounding control and the blue curve is the remaining confounding level after covariate balancing. The yellow shaded region (when the remaining confounding level exceeds the threshold) indicates that the MVTP shift may be too large to be balanced by the weights. The green curve indicates the largest $\Lambda$ for the sensitivity model to make the causal effect significant. A larger value indicates that the causal effect is more robust to unmeasured confounders. The causal effect is estimated over all ICU patients on the left panel and over only the ARDS patients on the right panel.}
    \label{fig:casestudy2}
\end{figure}

Figure \ref{fig:casestudy2} presents results for varying $\tau$ from 0.5 (reducing $V_T$ and $\textnormal{DP}$ but increasing $\textnormal{RR}$) to 1.5 (increasing $V_T$ and $\textnormal{DP}$ but decreasing $\textnormal{RR}$). The left panel of Figure \ref{fig:casestudy2} suggests that modifying driving pressure, while controlling for $\textnormal{C}{rs}$ and minute ventilation, does not significantly influence \textit{overall} patient mortality. However, when restricted to ARDS patients (535 out of 5011), the outcomes differ notably. As shown in the right panel, decreasing $\textnormal{DP}$ (with fixed $\textnormal{C}{rs}$ and minute ventilation) significantly improves survival among ARDS patients when $\tau < 0.85$, whereas increasing $\textnormal{DP}$ significantly worsens mortality. The black curve represents the 90\% permutation threshold for confounding, and the blue curve reflects the residual confounding after covariate balancing. These curves suggest our energy balancing approach effectively controls confounding within the $\tau$ range $[0.625, 1.5]$. For the largest significant $\Lambda$ value under the marginal sensitivity model, the green curve suggests that the estimated causal effect estimate is moderately robust with $\Lambda>1.5$ for shifts $\tau>1.25$. However, the beneficial effect of reducing driving pressure is less robust than the harmful effect of increasing it. These findings support the hypothesis that, for patients with ARDS, reducing driving pressure benefits patient mortality independently of patient severity, whereas increasing driving pressure harms patients with ARDS. This result is in agreement with the hypothesis that lung injury via DP is related to the elastance of the respiratory system.

\section{Discussion}
In this work, we aimed to provide analyses that give a better understanding of the factors that drive poor outcomes among ventilated patients in the ICU using observational data. 
To do so, our work leverages a modified vector-valued treatment policy (MVTP) approach to investigate complex causal relationships with multiple continuous-valued treatment variables, specifically addressing ventilator-induced lung injury (VILI). This approach not only facilitates robust causal inference but also provides clearer clinical interpretations by focusing on estimands with clear scientific meaning. We show that the MVTP framework transforms a vector-valued treatment problem into a simpler binary treatment problem, allowing for the use of the many available methods for the latter. By utilizing energy balancing tools that aim to balance covariate distributions, we provide highly robust confounding control in the presence of many confounders. We further make use of the connection between MVTPs and binary treatments by leveraging methods for sensitivity analysis to unmeasured confounding, a task critical for observational studies.

Through several analyses of the MIMIC-III dataset, our results reinforce the hypothesis suggested by \citet{marini2020component} that mechanical power, while sometimes used as an aggregate measure, may mask underlying nuances in how specific ventilation parameters contribute to lung injury. By separately evaluating modifications of tidal volume and airway pressures with our MVTP framework, we illustrate that equivalent reductions in mechanical power can have substantially different impacts on patient outcomes. Specifically, interventions aimed at reducing airway pressures—peak pressure, plateau pressure, and PEEP—appear to significantly decrease mortality risk, whereas similar proportional adjustments in tidal volume alone do not yield comparable survival benefits. This finding emphasizes the clinical relevance of individual ventilatory parameters beyond the general use of mechanical power, suggesting the importance of targeted ventilation strategies that prioritize the reduction of hazardous pressure thresholds.

Our secondary analysis of DP provides further clinical insights into its independent role in patient outcomes, particularly among ARDS patients. Previous literature indicates an association between lower DP and improved mortality; however, concerns persist regarding the potential confounding effect of respiratory-system compliance and disease severity. Our MVTP framework controls for these factors by proportionally adjusting tidal volume and respiratory rate alongside DP, effectively maintaining constant compliance and minute ventilation. In this context, we demonstrate a relationship between reduced DP and improved mortality specifically in ARDS patients. This strengthens the argument for DP-targeted ventilation strategies as a crucial component in lung-protective ventilation protocols, particularly among critically ill populations with compromised lung compliance.

\allowdisplaybreaks

\newenvironment{definition}[1][Definition]{\begin{trivlist}
		\item[\hskip \labelsep {\bfseries #1}]}{\end{trivlist}}
\newenvironment{example}[1][Example]{\begin{trivlist}
		\item[\hskip \labelsep {\bfseries #1}]}{\end{trivlist}}
\newenvironment{rmq}[1][Remark]{\begin{trivlist}
		\item[\hskip \labelsep {\bfseries #1}]}{\end{trivlist}}

\clearpage

\def\spacingset#1{\renewcommand{\baselinestretch}%
{#1}\small\normalsize} \spacingset{0.5}
\bibliographystyle{Chicago}
\bibliography{Bibliography}

\makeatletter\@input{xx_supp.tex}\makeatother

\end{document}

\if1\blind
{
	\title{\bf Supplementary material for ``Exploring the effects of mechanical ventilator settings with modified vector-valued treatment policies''}
	\author{Ziren Jiang$^{1}$, Philip S. Crooke$^{2}$, John J. Marini$^{3}$ Jared D. Huling$^{1}$\thanks{corresponding author: huling@umn.edu}\\ [8pt]
		$^{1}$Division of Biostatistics and Health Data Science,\\ University of Minnesota \\ [4pt]
        $^{2}$Department of Mathematics, Vanderbilt University \\ [4pt]
        $^{3}$Department of Pulmonary and Critical Care Medicine, \\ University of Minnesota  \\ [8pt]
	}
        \date{}
	\maketitle
} \fi

\if0\blind
{
	\title{\bf Supplementary material for ``Exploring the effects of mechanical ventilator settings with modified vector-valued treatment policies''}
	\date{}
	\maketitle
} \fi

%\pagenumbering{gobble}

\newpage
%\spacingset{1.75} % DON'T change the spacing!
\setstretch{1.775}
%\spacingset{1.5} % DON'T change the spacing!
\setlength{\abovedisplayskip}{7pt}%
\setlength{\belowdisplayskip}{7pt}%
\setlength{\abovedisplayshortskip}{5pt}%
\setlength{\belowdisplayshortskip}{5pt}%
%\pagenumbering{arabic}
%\setcounter{page}{1}
%%%%   ------------------------------------
%%%%     Input introduction section
%%%%   ------------------------------------
\section{Energy balancing tools for MVTP}
\subsection{Finite sample error decomposition for weighting estimators}

The idea of energy balancing weights comes from the decomposition of the estimation error for the weighted estimator $\frac{1}{n}\sum_{i=1}^n w_iY_i$ of the causal effect $\mu^q$ where $\{w_i\}_{i=1}^n$ are the balancing weights. Recall that $\mu(\x,\a)\equiv\mathbb{E}(Y|\X=\x,\A=\a)$ is the mean outcome function, the weighted estimator can be expressed as:
\begin{equation}\label{eq:weightedestimator}
    \hat{\mu}^{q}_{\w}=\frac{1}{n}\sum_{i=1}^{n} w_i Y_i=\frac{1}{n}\sum_{i=1}^{n} w_i\mu(\X_i,\A_i)+\frac{1}{n}\sum_{i=1}^{n}w_i\epsilon_i.
\end{equation}
where $\epsilon_i, i=1,...,n$ are the error term that have mean 0. Combining 
\begin{equation*}
    \mu^{q}=\int_{(\mathcal{X}, \mathcal{A})} \mu(\boldsymbol{x},\a) dF^{\q}_{\X,\A}(\boldsymbol{x},\a),
\end{equation*} 
with \eqref{eq:weightedestimator}, we have the following error decomposition for an arbitrary weighted estimator:
\begin{align}
       \hat{\mu}^{q}_{\w}-\mu^q={} & \underbrace{\int_{(\mathcal{X},\mathcal{A})}\mu(\boldsymbol{x},\a)d(F_{n,\w,\X,\A}-F_{n,\X,\A}^{q})(\boldsymbol{x},\a)}_{\text{error due to confounding}} \label{equ:errordecomp_bias}\\
        &+\underbrace{\int_{(\mathcal{X},\mathcal{A})}\mu(\boldsymbol{x},\a)d(F_{n,\X,\A}^{q}-F_{\X,\A}^{q})}_{\text{sampling error}} +\frac{1}{n}\sum_{i=1}^n w_i\epsilon_i, \label{equ:errordecomp_error}
\end{align}
where $F_{n,\X,\A}^{q}(\boldsymbol{x},\a)=\frac{1}{n}\sum_{i=1}^{n}\sum_{j=1}^{J(\X_i)} I_{j,\X_i}(\A_i)I(\X_i\leq \boldsymbol{x},q_j(\X_i,\A_i)\leq \a)$ is the empirical estimator of the shifted distribution $F_{\X,\A}^{q}$ (here the vector input indicator function $I(\X\leq\x)$ equals 1 only if $X_k\leq x_k$ is true for all dimension $k$); and $F_{n,\w,\X,\A}$ is the weighted empirical CDF with weights $\w$. From the error decomposition, since the second term is due to the sampling variability and the third term is of mean 0, the estimation bias can be mitigated by minimizing the first term which depends on the distance between the two distribution functions and the form of the outcome function. 

%\color{red}{Add another point of view of binary treatment}\color{black}

\subsection{Weighted energy distance}

The key component of the error decomposition \eqref{equ:errordecomp_bias} can be viewed as the integration of an outcome mean function over the difference between the empirical distribution of the observed sample and the target sample. Since the outcome mean function is arbitrary, minimizing the key component of error would require us to minimize the distributional difference. We adopt the energy distance \citep{szekely2013energy,huling2020energy} to measure the distributional difference that can be used in estimating the optimal balancing weights. 

The energy distance between the weighted empirical CDF $F_{n,\w,\X,\A}$ and the MVTP-shifted empirical CDF $F_{n,\X,\A}^{q}$ can be defined as:
\begin{align}\label{eq:energydistance}
&\mathcal{E}(F_{n,\w,\X,\A},F_{n,\X,\A}^{q}) \nonumber \\ 
&= {}  \frac{1}{n^2}\bigg\{2\times\sum_{i=1}^n\sum_{k=1}^n\sum_{j=1}^{J(\X_k)}I_{j,\X_k}(\A_k)w_i||(\X_i,\A_i)-(\X_k,q_j(\X_k,\A_k))||_2 \nonumber\\
        &\quad\quad-\sum_{i=1}^n\sum_{k=1}^n w_i w_k ||(\X_i,\A_i)-(\X_k,\A_k)||_2 \nonumber\\
        &\quad\quad-\sum_{i=1}^n\sum_{j=1}^{J(\X_i)}\sum_{k=1}^n\sum_{j'=1}^{J(\X_k)} I_{j,\X_i}(\A_i) I_{j',\X_k}(\A_k)||(\X_i,q_j(\X_i,\A_i))-(\X_k,q_{j'}(\X_k,\A_k))||_2 \bigg\},
\end{align}
which measures the distributional difference between the two populations.

\subsection{Energy balancing weights}

Since the finite sample error decomposition \eqref{equ:errordecomp_bias} indicates that the estimation bias of the causal effect can be eliminated by minimizing the empirical distribution of the observed population and the target population, which can be measured by their energy distance, we can find balancing weights that directly minimize the weighted energy distance \eqref{eq:energydistance}.

We define the penalized energy balancing weights $\w^{p}_n$ with a user-specified parameter $\lambda>0$ as:
\begin{equation}\label{equ:optipen}
         \begin{split}
        \w_n^{p}&\in \underset{\w=(w_1,...,w_n)}{\arg\min} \mathcal{E}(F_{n,\w,\X,A},F_{n,\X,\A}^{q})+\frac{\lambda}{n^2}\sum_{i=1}^nw_i^2 \textnormal{ s.t. } \sum_{i=1}^n w_i=n\,, w_i\geq0.
    \end{split}
\end{equation}
The corresponding penalized energy balancing estimator is then $\hat{\mu}^{q}_{\w^{p}_n}=n^{-1}\sum_{i=1}^n \w_{i,n}^{p}Y_i$. Under mild conditions, \citet{jiang2023enhancing} showed that: (1) the penalized energy balancing weights make the weighted empirical CDF of the observed data converge to the true MVTP-shifted CDF, and most importantly (2) the penalized energy balancing estimator achieves root-$n$ consistency.

We next introduce the augmented energy balancing estimator which incorporates a fitted outcome regression model. Let $\hat{\mu}(\boldsymbol{x},\a)$ be an estimate of the outcome regression function $\mu(\boldsymbol{x},\a)$, then the augmented energy balancing estimator is defined as:
\begin{equation}
\begin{split}
        \hat{\mu}_{AG}^{q}&=\hat{\mu}_{\w_n^{p}}^{q}-\int_{\mathcal{X}}\int_{\mathcal{A}}\hat{\mu}(\boldsymbol{x},\a)d(F^{q}_{n,\X,\A}-F_{n,w^{p}_n,\X,\A})(\boldsymbol{x},\a)\\
        &=\frac{1}{n}\sum_{i=1}^nw^{p}_i(Y_i-\hat{\mu}(\X_i,\A_i))+\frac{1}{n}\sum_{i=1}^n\sum_{j=1}^{J(\X_i)}I_{j,\X_i}(\A_i)\hat{\mu}(\X_i,q_j(\X_i,\A_i)).
\end{split}
\end{equation}
The augmented energy balancing estimator is composed of an outcome prediction term and a weighted residual term which can be viewed as a bias correction based on the balancing weights. \citet{jiang2023enhancing} further showed its asymptotic normality under certain regularization conditions.

\subsection{Selection of a feasible MVTP scale}

One major challenge for causal inference with continuous treatment is the validity of the positivity assumption. This can be a more serious problem for the vector-value treatment due to the additional complexity of multiple continuous treatment variables. Therefore, it is important to make sure the validity of the positivity assumption before interpreting the estimated causal effect for any particular MVTP. Our second tool for causal inference of MVTP is a permutation test for testing whether there exist imbalance after applying the balancing weights. 

For this test, the null hypothesis is that the weighted population is identical to the target population (i.e., the MVTP-shifted population $F^q_{\X,\A}$). Under the null hypothesis, the energy distance between the two empirical populations ($F_{n, \w, \X,\A}$ and $F^q_{n, \X,\A}$) should converge to 0 when $n$ goes to infinite; and it is non-zero only due to the sampling variability for finite samples. We use the permutation method to estimate the upper threshold (critical value) of the sampling variability under the null hypothesis. As illustrated in Figures 4 and 5 of the main manuscript, if the weighted energy distance exceeds this threshold, we may reject the null hypothesis and conclude that the balancing weights can not effectively balance the measured confoundings due to the potential positivity violation. For a detailed description of this permutation test, readers are referred to \citet{jiang2023enhancing} Section 4.1.

\section{Supplementary simulation study materials}
\subsection{Data-generating mechanism}\label{sec:datagene}
This subsection contains full details of the data-generating mechanisms used in the main text. They are described as follows:
\begin{itemize}
    \item The covariates and treatment variables are generated follows the plasmode simulation technique described in the main manuscript.
    \item Given the covariates $(X_{i1},...,X_{ip})$ and treatment $(A_{i1}, A_{i2}, A_{i3}, A_{i4}, A_{i5})$, generate the outcome $Y_i\sim N(\mu_i,1)$ where the mean $\mu_i$ follows,
    \begin{equation*}
        \begin{split}
            &\mu_i=\bigg(-1-\frac{1}{2}X_{i3}+\frac{1}{2}X_{i4}^2+\frac{1}{2}X_{i5}^2+\sum_{j \text{ is even},j=12}^p (\frac{3}{2}X_{ij}+X_{ij}(X_{i(j-10)}+2X_{i(j-9)}))\bigg) \\
            &\times\bigg(\frac{0.0005}{(p+20)^2}(A_{i1}+A_{i2})\bigg)+
            \\
            &  \bigg(1-\frac{1}{2}X_{i1}^2-\frac{1}{2}X_{i2}^2+X_{i1}X_{i2}+\frac{1}{2}(X_{i7}+X_{i8}+X_{i10})-\\
            & \sum_{j \text{ is odd},j=11}^p (-\frac{1}{2}X_{ij}+0.3X_{ij}^2+0.3X_{ij}(X_{i(j-8)}^2+2X_{i(j-7)}))\bigg)^2 \\
            & \times \bigg((\frac{A_{i3}-A_{i4}+\frac{1}{2}A_{i5}-20}{2})^2-6\bigg)\times 20
        \end{split}
    \end{equation*}
    \item The modified treatment policy is $q(\x,\a)$ designed to  depend on the original treatment variable $\a$ and the modificationmagnitude $\tau$, 
\begin{equation*}
     q(\x,\a) = (1-\tau)\a
\end{equation*}
\end{itemize}

\section{Proof of the theorems}
\subsection{Proposition 1}
\begin{prop}
    Under assumptions \textbf{A0}-\textbf{A2} and condition C1,
    \begin{equation}\label{eq:identification}
    \mu^{q}=\int_{\mathcal{X}} \sum_{j=1}^{J(\x)} \int_{h_j(\x,\a)\in I_{j,\x}} \mathbb{E}(Y|\X=\boldsymbol{x},\A=\a)dF_{\X,\A}(\boldsymbol{x},\h_j(\boldsymbol{x},\a)).
\end{equation}
\end{prop}

\begin{proof}
    By definition, we have:
    \begin{equation*}
        \begin{split}
            \mu^{q}&\equiv\int_{(\mathcal{X},\mathcal{A})} \mathbb{E}[Y^{\q(\boldsymbol{x},\a)}|\X=\boldsymbol{x},\A=\a]dF_{\X,A}(\boldsymbol{x},\a)\\
            &=\int_{\mathcal{X}} \sum_{j=1}^{J(\x)}\int_{\a\in I_{j,\x}} \bbE(Y^{\q_j(\x,\a)}|\X=x,\A=\a) dF_{\X,\A}(\x,\a).
        \end{split}
    \end{equation*}
    By the definition of the integral, we can substitute the variable $\a$ with $h_j(\x,\a)$ for each $\int_{\a\in I_j(\x)} E(Y^{\q_j(\x,\a)}|\X=\x, \A=\a)dF(\x,\a)$ where $h_j(\x,\cdot)$ be the inverse function of $q_j(\x,\cdot)$. Thus, we have:
    \begin{equation*}
        \begin{split}
            \mu^{q}&=\int_{\mathcal{X}} \sum_{j=1}^{J(\x)}\int_{\a\in I_{j,\x}} \bbE(Y^{\q_j(\x,\a)}|\X=\x,\A=\a) dF_{\X,\A}(\x,\a)\\
            &=\int_{\mathcal{X}} \sum_{j=1}^{J(\x)}\int_{\h_j(\x,a)\in I_{j,\x}} \bbE(Y^{\q_j(\x,\h_j(\x,\a))}|\X=\x,\A=\h_j(\x,\a)) dF_{\X,\A}(\x,\h_j(\x,\a))\\
            &=\int_{\mathcal{X}} \sum_{j=1}^{J(\x)}\int_{\h_j(\x,\a)\in I_{j,\x}} \bbE(Y^{\a}|\X=\x,\A=\h_j(\x,\a)) dF_{\X,\A}(\x,\h_j(\x,\a))\\
            &=\int_{\mathcal{X}} \sum_{j=1}^{J(\x)}\int_{\h_j(\x,\a)\in I_{j,\x}} \bbE(Y^{\a}|\X=\x,\A=\a) dF_{\X,\A}(\x,\h_j(\x,\a))\\
            &=\int_{\mathcal{X}} \sum_{j=1}^{J(\x)}\int_{\h_j(\x,\a)\in I_{j,\x}} \bbE(Y|\X=\x,\A=\a) dF_{\X,\A}(\x,\h_j(\x,\a)),
        \end{split}
    \end{equation*}
    where the third equality holds because $\q_j(\x,\h_j(\x,\a))=\a$. The second to last equation is due to the conditional exchangeability of the related population assumption (\textbf{A2}) which states that $Y^{\a}|\X=\x,\A=\a$ and $Y^{\a}|\X=\x,\A=\a'$ have the same distribution as long as $\a=\q(\x, \a')$. The last equality is valid due to the consistency assumption (\textbf{A0}).
\end{proof}

\subsection{Lemma 2}
\begin{lemma}\textnormal{(Equivalence of the causal effect between the MVTP and binary treatment)} For the augmented population $\{\widetilde{\X},\widetilde{\A}, Y, Z\}\equiv\{\X, \A, Y, Z = 0\}\cup\{\X, \q(\X, \A), Y, Z = 1\}$ defined in Section 3.2, consider $Z$ as the binary treatment and $(\X, \A)$ as the covariates. Then the average treatment effect on the treated (ATT), under the common consistency, positivity, and ignorability assumptions for binary treatment, equals the causal effect of MVTP under assumption \textnormal{\textbf{A0}} - \textnormal{\textbf{A2}}:
\begin{equation}
    \mathbb{E}(Y(1)-Y(0)|Z=1) = \mu^q-\mathbb{E}(Y).
\end{equation}
Furthermore, the positivity and ignorability assumptions for binary treatment are equivalent to assumptions \textbf{A1} and \textbf{A2} for MVTP.

\end{lemma}
\begin{proof}
    \begin{equation}
    \begin{split}
        \mathbb{E}(Y(1)-Y(0)|Z=1) &= \mathbb{E}(Y(1)|Z=1)-\mathbb{E}(Y(0)|Z=1)\\
        \textnormal{(consistency)}&= \mathbb{E}(Y|Z=1) - \mathbb{E}(Y(0)|Z=1)\\
        \textnormal{(double expectation)}&= \mathbb{E}(Y) - \mathbb{E}(\mathbb{E}(Y(0)|\X,\A,Z=1)|Z=1)\\
        \textnormal{(ignorability)}&= \mathbb{E}(Y) -\mathbb{E}(\mathbb{E}(Y(0)|\X,\A,Z=0)|Z=1)\\
        \textnormal{(consistency)}&= \mathbb{E}(Y) -\mathbb{E}(\mathbb{E}(Y|\X,\A,Z=0)|Z=1)\\
        \textnormal{(definition of $\mu$)}&= \mathbb{E}(Y) - \mathbb{E}(\mu(\X,\A)|Z=1)\\
        &=\mathbb{E}(Y) - \int_{(\mathcal{X}, \mathcal{A})} \mu(\boldsymbol{x},\a) dF^{\q}_{\X,\A}(\boldsymbol{x},\a) \\
        &= \mathbb{E}(Y) - \mu^q.
    \end{split}
    \end{equation}

    For the equivalence of the positivity assumption for binary treatment $Z$ to assumption \textbf{A1}, note that the positivity assumption for ATT requires that the support of covariates $(\X,\A)$ in the $Z=0$ population contains the support of $(\X, \q(\X,\A))$ in the $Z=1$ population. Since the support of $(\X, \q(\X,\A))$ in the $Z=1$ population is defined as $\{(\x,\q(\x,\a): (\x,\a)\in (\mathcal{X}, \mathcal{A}))\}$, the positivity assumption is equivalent to assume $(\x,\q(\x,\a))\in (\mathcal{X}, \mathcal{A})$ for any $(\x,\a)\in (\mathcal{X}, \mathcal{A})$, which is exactly the assumption \textbf{A1} for MVTP. 
\end{proof}

\subsection{Lemma 3}

\begin{lemma}\textnormal{(Equivalence between the assumption \textbf{A2} and the ignorability assumption)} 
For the augmented population $\{\widetilde{\X},\widetilde{\A}, Y, Z\}\equiv\{\X, \A, Y, Z = 0\}\cup\{\X, \q(\X, \A), Y, Z = 1\}$ defined in Section 3.2, consider $Z$ as the binary treatment and $(\X, \A)$ as the covariates. Then the ignorability assumption for ATT is equivalent to the conditional exchangeability of related population assumption \textbf{A2} for the MVTP. 

\end{lemma}

\begin{proof}
     The ignorability assumption for ATT requires that $Y(0)$ is independent of $Z$ given the covariates $\X=\x,\A=\a$. In other words, it requires that $Y(0)|Z=0,\X=\x,\A=\a$ and $Y(0)|Z=1,\X=\x,\A=\a$ have the same distribution. Since $Y(0)$ is the outcome variable $Y$ in the observed population, the ignorability assumption for ATT requires that we include all covariates such that it simultaneously (1) influences the outcome $Y$, and (2) it has a different distribution in the population of $Z=1$ and population of $Z=0$. To see how this connects with the conditional exchangeability assumption \textbf{A2}, recall that \textbf{A2} requires $Y^{\a'}|\X=\x,\A=\a$ and $Y^{\a'}|\X=\x,\A=\a'$ have the same distribution for each $(\x,\a)\in(\mathcal{X},\mathcal{A})$. The only scenario that violates this assumption is the existence of an unmeasured variable such that (1) it influences the outcome $Y$, and (2) it is imbalanced in the observed population $(\X,\A)$ and the MVTP shifted population $(\X, \q(\X,\A))$. Since the $Z=0$ population is exactly the observed population and the $Z=1$ population is the MVTP shifted population, the equivalence is obvious. 
\end{proof}

\subsection{Lemma 4}
\begin{lemma}\textnormal{(Connection between the Marginal sensitivity model and ratio of the weights)} Denote the true data-generating mechanism as $\mathbb{P}(\X,\A,Y,\U)$ where $\U$ is the unobserved confounders such that the conditional exchangeability of related populations assumption \textbf{A2} hold with the observe of $\U$. Then for $\Lambda\geq 1$, the marginal sensitivity model \textbf{A3} is equivalent to assume the true data-generating mechanism $\mathbb{P}(\X,\A,Y,\U)$ follows:
\begin{equation}
    \mathbb{P}(\X,\A,Y,\U)\in\mathcal{P}(\Lambda)=\{\mathbb{P}(\X,\A,Y,\U):\Lambda^{-1}\leq  \frac{w}{w^*} \leq\Lambda \}
\end{equation}
where $w$ is the density ratio weights under the observation of $(\X, \A)$ and $w^*$ is the ``true" density ratio weights if we observe $(\X, \A, \U)$.
\end{lemma}
\begin{proof}
    Under the observation of $(\X, \A)$, the density ratio weights $w$ can be expressed as 
    \begin{equation*}
        w(\x,\a)=\frac{d F^q_{\X,\A}(\x,\a)}{d F_{\X,\A}(\x,\a)} = \frac{\pi(\x,\a)}{1-\pi(\x,\a)}
    \end{equation*}
    Similarily, under the observation of the full data $(\X, \A, \U)$, the density ratio weights $w^*$ can be expressed using the true propensity score $\pi^*(\x,\a,y)$:
    \begin{equation*}
        w^*(\x,\a,y)=\frac{\pi^*(\x,\a,y)}{1-\pi^*(\x,\a,y)}
    \end{equation*}
    Then the odds ratio of the two propensity scores $\textnormal{OR}(\pi(\x,\a),\pi^*(\x,\a,y))$ becomes 
    \begin{equation*}
        \textnormal{OR}(\pi(\x,\a),\pi^*(\x,\a,y))=\frac{\pi(\x,\a)/(1-\pi(\x,\a))}{\pi^*(\x,\a,y)/(1-\pi^*(\x,\a,y))}=\frac{w}{w^*}
    \end{equation*}
    which finishes the proof.
\end{proof}

\allowdisplaybreaks

\newenvironment{definition}[1][Definition]{\begin{trivlist}
		\item[\hskip \labelsep {\bfseries #1}]}{\end{trivlist}}
\newenvironment{example}[1][Example]{\begin{trivlist}
		\item[\hskip \labelsep {\bfseries #1}]}{\end{trivlist}}
\newenvironment{rmq}[1][Remark]{\begin{trivlist}
		\item[\hskip \labelsep {\bfseries #1}]}{\end{trivlist}}

\clearpage

\bibliographystyle{Chicago}
\bibliography{Bibliography}

\makeatletter\@input{xx.tex}\makeatother